\documentclass[3p,times,twocolumn,procedia,sort&compress]{elsarticle}

\usepackage{ecrc}

\volume{00}
\firstpage{1}
\journalname{Nuclear Physics B Proceedings Supplement}
\runauth{J. Berges et al.}

\jid{nuphbp}

\CopyrightLine{2012}{Published by Elsevier Ltd.}

\usepackage[figuresright]{rotating}

\usepackage{axodraw4j}
\usepackage{amsmath,amssymb}
\usepackage{dsfont}
\usepackage[retainorgcmds]{IEEEtrantools}
\usepackage{picture}
\usepackage{graphicx}
\usepackage{ebezier}
\usepackage{color}
\usepackage{multirow,bigdelim,booktabs}

\DeclareMathOperator{\sgn}{sgn}
\DeclareMathOperator{\Tr}{Tr}
\newcommand\ud{d}
\newcommand\rmR{\mathrm R}
\newcommand\rmA{\mathrm A}
\newcommand\eqn[1]{\begin{equation*} #1 \end{equation*}}
\newcommand\eqnlabel[1]{\begin{equation} #1 \end{equation}}
\newcommand\eqna[1]{\begin{IEEEeqnarray*}{rCl} #1 \end{IEEEeqnarray*}} 
\newcommand\mbf[1]{\boldsymbol{\mathit{#1}}}
\newcommand\avr[1]{\left\langle{#1}\right\rangle}
\newcommand\lbr[1]{\langle{#1}\,|}
\newcommand\rbr[1]{|\,{#1}\rangle}
\newlength{\pmsk}
\begin{document}

\def\ColdAtoms{Berloff:2002,Scheppach:2009wu,Nowak:2011sk}
\def\RGReviews{Morris:1998da,Aoki:2000wm,Bagnuls:2000ae,Berges:2000ew,Polonyi:2001se,Salmhofer:2001tr,Pawlowski:2005xe,Gies:2006wv,Delamotte:2007pf,Rosten:2010vm,Kopietz:2010zz}

\begin{frontmatter}

\title{Introduction to the nonequilibrium\\ functional renormalization group}

\author{J.~Berges}
\author{D.~Mesterh\'azy}

\address{Institut f\"ur Theoretische Physik, Universit\"at Heidelberg, Philosophenweg 16, 69120 Heidelberg, Germany}

\begin{abstract}
In these lectures we introduce the functional renormalization group out of equilibrium. While in thermal equilibrium typically a Euclidean formulation is adequate, nonequilibrium properties require real-time descriptions. 
For quantum systems specified by a given density matrix at initial time, a generating functional for real-time correlation functions can be written down using the Schwinger-Keldysh closed time path. This can be used to construct a nonequilibrium functional renormalization group along similar lines as for Euclidean field theories in thermal equilibrium. Important differences include the absence of a fluctuation-dissipation relation for general out-of-equilibrium situations. The nonequilibrium renormalization group takes on a particularly simple form at a fixed point, where the corresponding scale-invariant system becomes independent of the details of the initial density matrix. We discuss some basic examples, for which we derive a hierarchy of fixed point solutions with increasing complexity from vacuum and thermal equilibrium to nonequilibrium. The latter solutions are then associated to the phenomenon of turbulence in quantum field theory.
\end{abstract}


\end{frontmatter}

\hspace*{2.cm}

\section{Introduction}
\label{sec:Intro}

Thermal equilibrium properties of many-body or field theories are known to be efficiently classified in terms of renormalization group fixed points. A particularly powerful concept is the notion of infrared fixed points which are characterized by universality. These correspond to critical phenomena in thermal equilibrium, where the presence of a characteristic large correlation length leads to independence of long-distance properties from details of the underlying microscopic theory. In contrast, a classification of properties of theories far from thermal equilibrium in terms of renormalization group fixed points is much less developed.

\hspace*{2.cm}

\begin{eqnarray}
\nonumber
\end{eqnarray}

The notion of universality or criticality far from equilibrium is to a large extent unexplored, in particular, in relativistic quantum field theories. Here, the strong interest is mainly driven by theoretical and experimental advances in our understanding of early-universe cosmology as well as relativistic collision experiments of heavy nuclei in the laboratory.

In the latter contexts, a particular class of nonthermal fixed points has attracted much interest in recent years. It is associated to the phenomenon of turbulence in quantum field 
theory, where a universal power-law behavior describes the transport of conserved quantities. Traditionally, turbulence is associated mostly with the dynamics of vortices in fluids \cite{Frisch:318247} but also nonlinear waves can show turbulent behavior \cite{Nazarenko:1399206}. In particular, it is well-known that interacting quantum field theories can lead to nonlinear dynamics and wave turbulence, even for very weakly coupled theories. 

Among the best studied theoretical examples in relativistic quantum field theory are scalar inflaton models for the dynamics of the early universe \cite{Kofman:2008zz}. In a large class of models, the strongly accelerated expansion of the universe after the Big Bang is followed by turbulent behavior of the inflaton field before 
thermal equilibrium is achieved. The connection to the well-established phenomenon of weak wave turbulence in the presence of small nonlinearities has been studied in great detail 
\cite{Micha:2004bv}. Here, weak wave turbulence is associated to an energy cascade from small to high wave numbers. Only recently it has been realized that the direct energy transport 
towards higher wavenumbers is part of a dual cascade, in which also an inverse particle flux towards the infrared at small wave numbers occurs \cite{Berges:2008wm}. One of the striking 
consequences is Bose condensation far from equilibrium \cite{Berges:2012us}. Similar scaling phenomena may also occur for gauge field dynamics in the context of heavy-ion 
collisions at sufficiently high energies \cite{Berges:2008mr,Carrington:2010sz,Fukushima:2011nq,Berges:2012ev}, or also for the nonrelativistic dynamics of ultracold atoms \cite{\ColdAtoms}. The emergence of same macroscopic scaling phenomena from very different underlying microscopic physics is a formidable manifestation of universality far from equilibrium.
   
Understanding the dominant collective phenomena in quantum field theories far from equilibrium represents a major challenge. Important phenomena, such as the infrared particle cascade and subsequent Bose condensation mentioned above, are genuinely nonperturbative and require suitable approximation techniques. Here a nonequilibrium functional renormalization group approach \cite{Canet:2003yu,Gezzi:2006,Jakobs:2007,Gasenzer:2008zz,Berges:2008sr,Gasenzer:2010rq,Kloss:2010wj,Canet:2011wf}, or related real-time functional integral techniques based on $n$-particle irreducible (nPI) effective actions \cite{Berges:2004yj}, can serve as a very useful means to gain analytic understanding. Other implementations of the renormalization group idea in this context include the 
so-called numerical renormalization group approach \cite{Anders:2005}, the time-dependent density matrix renormalization group \cite{Schmitteckert:2004}, the real-time renormalization group in Liouville space \cite{Schoeller:1999ed}, or flow equations describing infinitesimal unitary transformations \cite{Kehrein:2005}. These approaches can be complemented by numerical simulations in (classical-statistical) nonequilibrium lattice theories \cite{Prokopec:1996rr,Berges:2005yt,Duben:2008iw,Mesterhazy:2011kr} or using kinetic descriptions \cite{Blaizot:2001nr} in their respective range of applicability. 

In these lectures we discuss basic properties of the nonequilibrium renormalization group for the scale dependent generating functional of 1PI correlation functions in relativistic quantum field theory, following closely Ref.~\cite{Berges:2008sr}. Concentrating on the explicit example of a $N$-component scalar field theory allows us to focus on the relevant differences to Euclidean treatments in vacuum or thermal equilibrium, without introducing too much formalism and abstract notation. With the help of standard references from the functional renormalization group in Euclidean spacetime \cite{\RGReviews}, one can apply these considerations in a similar way to include fermions or gauge fields. 

We begin in section \ref{sec:basics} by emphasizing some important differences between thermal equilibrium and nonequilibrium. This will help us to understand why there exists a hierarchy of fixed point solutions with increasing complexity from vacuum and thermal equilibrium to nonequilibrium. In section \ref{sec:weakturbulence} we introduce the notion of scaling behavior far from equilibrium in the context of weak wave turbulence for the description of stationary transport of conserved quantities. The nonequilibrium functional renormalization group is introduced in section \ref{sec:NEQFRG}, where we derive the relevant flow equations. In section \ref{sec:truncation} we solve the flow equations in an approximation based on a resummed large-$N$ expansion to next-to-leading order. The results are used in section \ref{sec:nonthermalfp} to determine scaling exponents for nonthermal fixed points. We summarize and give an outlook in section \ref{sec:summary}.

\section{Thermal equilibrium vs.\ non\-equi\-lib\-rium}
\label{sec:basics}

\subsection{Statistical and spectral functions}
\vskip\pmsk

All information about a quantum theory is encoded in its correlation functions for a given density matrix $\varrho$. In thermal equilibrium, for the case of a canonical ensemble with inverse temperature $\beta \equiv 1/T$ and Hamilton operator $H$, the density matrix is given by $\varrho^{\rm (eq)} \sim e^{-\beta H}$ and it is normalized such that $\Tr \varrho^{\rm (eq)} = 1$.
A thermal real-time correlation function for a Heisenberg field operator $\Phi(x)$ is given by the time-ordered trace
\eqna{
G^{\rm (eq)}(x-y) &=& \Tr \left\{\varrho^{\rm (eq)} \,\textrm{T}_0 \Phi(x) \Phi(y) \right\} \\
&=& \langle \Phi(x) \Phi(y)\rangle_{\rm eq}\, \theta(x^0-y^0)  \\
&& +\: \langle \Phi(y) \Phi(x)\rangle_{\rm eq}\, \theta(y^0-x^0) ~,
\label{eq:thermalG} \IEEEyesnumber}
for a two-point function, and involves $n$ fields for a $n$-point function. Here, $\textrm{T}_0$ is the time-ordering operator, $x=(x^0,\mbf{x})$ denotes the time $x^0$ and space $\mbf{x}$ variables\footnote{We use a metric with signature $(+,-,-,-)$.} and, to be specific, we consider  real scalar fields.

Nonequilibrium typically requires the specification of a density matrix $\varrho(t_0)$ at some initial time $t_0$ where $\varrho(t_0) \neq \varrho^{\rm (eq)}$. The task of nonequilibrium quantum field theory is then to determine the real-time evolution of correlation functions such as the two-point function 
\eqnlabel{
G(x,y) = \Tr\avr{\varrho(t_0) \,\textrm{T}_0 \Phi(x) \Phi(y)}
\equiv \avr{ \textrm{T}_0 \Phi(x) \Phi(y)} ~,
\label{eq:twopoint}}
for times $x^0, y^0 > t_0$. In general, translational invariance does not hold out of equilibrium and thus, two-point functions will depend on both $x$ and $y$ separately, i.e.
\eqnlabel{G(x,y) 
= F(x,y) - \frac{i}{2} \rho(x,y) \sgn(x^{0} - y^{0}) ~.
\label{2ptFunction}}
In the second line of \eqref{2ptFunction} we introduced the statistical two-point function $F(x,y)$ and the spectral function $\rho(x,y)$. These are given by the expectation values of the anti-commutator of the scalar field\footnote{In this section, we will assume for simplicity that we are in the symmetric phase where the field expectation value $\phi(x) \equiv \avr{\Phi(x)}$ vanishes. Otherwise the connected statistical function is given by $F(x,y) = \frac{1}{2} \avr{\{ \Phi(x), \Phi(y) \}} - \phi(x) \phi(y)$.}
\eqnlabel{F(x,y) = \frac{1}{2} \avr{\{ \Phi(x), \Phi(y) \}} ~,
\label{StatisticalFunction}}
and the commutator
\eqnlabel{\rho(x,y) = i \avr{[ \Phi(x) , \Phi(y)]} ~,
\label{SpectralFunction}}
respectively. The decomposition \eqref{2ptFunction} follows from \eqref{eq:thermalG} with the elementary properties of the Heaviside step function, $\theta(x^{0} - y^{0}) + \theta(y^{0} - x^{0}) = 1$ and $\sgn(x^{0} - y^{0}) = \theta(x^{0} - y^{0}) - \theta(y^{0} - x^{0})$. The spectral function is also related to the retarded and advanced Green's function by $G^R(x,y) = \theta(x^0-y^0) \rho(x,y) = G^A(y,x)$, respectively. 
Loosely speaking, the spectral function $\rho(x,y)$ determines which states are available while the statistical function $F(x,y)$ contains the information about how often a state is occupied \cite{Berges:2004yj}.

\subsection{Absence of a fluctuation-dissipation relation}
\vskip\pmsk

In contrast to the general nonequilibrium case, it is an important simplification of thermal or vacuum theories that the statistical \eqref{StatisticalFunction} and spectral function \eqref{SpectralFunction} are related by the so-called fluctuation-dissipation relation. We discuss this relation for the two-point correlation function with a canonical density matrix $\varrho^{\rm (eq)} \sim e^{-\beta H}$. For times $t > 0$ the thermal two-point function \eqref{eq:thermalG} reads 
\eqna{
&& \hspace{-30pt} \Tr\left\{ e^{-\beta H} \Phi(t,\mbf{x}) \Phi(0,\mbf{y}) \right\} \\
&=& \Tr\Big\{ e^{-\beta H} \underbrace{e^{\beta H} \Phi(0,\mbf{y}) e^{-\beta H}}_{= \,\Phi(-i \beta,\mbf{y})}  \Phi(t,\mbf{x}) \Big\} 
\label{eq:step1} \IEEEyesnumber ~. 
}
In  the second line we have used the invariance of the trace under cyclic changes and inserted $e^{-\beta H}e^{\beta H} = 1$. From the real-time Heisenberg evolution $\Phi(t,\mbf{x}) = e^{i H t} \Phi(0,\mbf{x}) e^{- i H t}$, we may \emph{define} in complete analogy $\Phi(-i \beta,\mbf{x}) \equiv e^{\beta H} \Phi(0,\mbf{x}) e^{- \beta H}$ as an extension to `imaginary times' \cite{Landsman:1986uw}. We can state the above relation directly in terms of the statistical \eqref{StatisticalFunction} and spectral function \eqref{SpectralFunction} using $\langle\Phi(t,\mbf{x}) \Phi(0,\mbf{y}) \rangle = 
F(t,0;\mbf{x},\mbf{y}) - \frac{i}{2} \rho(t,0;\mbf{x},\mbf{y})$, and correspondingly for the r.h.s.\ of \eqref{eq:step1}. Since equilibrium is spacetime translation invariant, \eqref{eq:step1} then reads
\eqna{
&& \hspace{-20pt} \left( F^{\rm (eq)}(x-y) - \frac{i}{2}\rho^{\rm (eq)}(x-y) \right)\Big|_{y^{0} = 0}
\\ 
&=& \left( F^{\rm (eq)}(x-y) + \frac{i}{2}\rho^{\rm (eq)}(x-y) \right)\Big|_{y^{0} = -i\beta}
\label{KMS}\IEEEyesnumber ~ .}
Translation invariance also makes it convenient to consider the Fourier transform with {\it real} four-momentum $p=(p^0,\mbf{p})$, that is
\eqnlabel{F^{\rm (eq)}(x-y) = \int\!\!\frac{d^{d+1} p}{(2\pi)^{d+1}} e^{-i p (x-y)} F^{\rm (eq)}(p) ~,
\label{eq:realFourier} \IEEEyesnumber}
and equivalently for $\rho^{\rm (eq)}(x-y)$. Taking for a moment for granted that these integrals can be properly regularized and defined for the considered quantum field theory, we obtain from \eqref{KMS}
\eqna{&& \hspace{-20pt} F^{\rm (eq)}(p) - \frac{i}{2} \rho^{\rm (eq)} (p) \\ &=& e^{\beta p^{0} } \left( F^{\rm (eq)}(p) + \frac{i}{2} \rho^{\rm (eq)} (p) \right) ~.}
This can be written in the form of a fluctuation-dissipation relation as
\eqnlabel{F^{\rm (eq)}(p) = -i \left( n_{\rm BE}(p^{0}) + \frac{1}{2} \right) \rho^{\rm (eq)} (p) ~,
\label{FDR}}
where 
\eqnlabel{n_{\rm BE}(p^{0}) = \frac{1}{e^{\beta p^{0}} - 1} ~,
\label{eq:BE}}
is the Bose-Einstein distribution function. Since this distribution depends on frequency $p^0$ and \emph{not} on spatial momenta $\mbf{p}$, it relates the anti-commutator expectation value of fields ($F$) and the respective commutator ($\rho$) in a nontrivial way. Moreover, at zero temperature the distribution function $n_{BE}$ is zero and the anti-commutator and commutator are directly proportional to each other. In contrast, for a nonequilibrium density matrix no such constraints exist in general. As a consequence, $F$ and $\rho$ are linearly independent out of equilibrium. In particular, the absence of a fluctuation-dissipation relation will allow us to observe new types of scaling solutions, beyond those known from thermal equilibrium or the vacuum.

\subsection{Infrared fixed points}
\vskip\pmsk

Renormalization group fixed points correspond to scaling solutions for correlation functions. Before computing them from first principles in the main part of these lectures, we illustrate here heuristically possible scaling behaviors in and out of equilibrium. For that purpose, we consider translationally invariant spectral and statistical two-point functions, $F(x-y)$ and $\rho(x-y)$, assuming also spatial isotropy to limit the number of possible scaling exponents. The Fourier transforms, $F(p^0,\mbf{p})$ and $\rho(p^0,\mbf{p})$, then depend on real frequency and momenta, where any scaling \emph{ansatz} has to take into account the different possible scalings of spatial momenta vs.\ frequencies. This difference is described by the `dynamical critical exponent' $z$. Furthermore, we denote the overall scaling of the statistical two-point function by the `occupation number exponent' $\kappa$ and of the spectral function by the `anomalous dimension' $\eta$. The scaling behavior may then be described as        
\eqna{F(p^{0},\mbf{p}) &=& s^{2 + \kappa} F(s^{z} p^{0} , s \mbf{p}) ~, \IEEEyessubnumber \label{ScalingAnsatz1}\\
\rho(p^{0} , \mbf{p}) &=& s^{2- \eta} \rho(s^{z} p^{0}, s \mbf{p}) ~, \IEEEyessubnumber \label{ScalingAnsatz2}}
for any real scaling parameter $s > 0$.

Before we consider such a scaling behavior for the nonequilibrium case, it is very instructive to first insert \eqref{ScalingAnsatz1} and \eqref{ScalingAnsatz2} into the fluctuation-dissipation relation \eqref{FDR} in order to see how this constraint relates the different exponents. We restrict the discussion to frequencies (and momenta) much smaller than the temperature $T$. This would be the relevant range for scaling behavior, for instance, near second-order phase transitions in thermal equilibrium. In the infrared, where $p^{0} \ll T$, we see that the distribution function \eqref{eq:BE} assumes the scaling form
\eqnlabel{n_{\rm BE} (p^{0}) \sim \frac{T}{p^{0}} 
\label{eq:classicaldist}~.}
In particular, occupation numbers become large such that the `proportionality factor' $n_{\rm BE}(p^0) + 1/2$ appearing in \eqref{FDR} can always be replaced by \eqref{eq:classicaldist} for sufficiently small $p^0$. 
Using \eqref{ScalingAnsatz1} and \eqref{ScalingAnsatz2}
we can then directly read off from \eqref{FDR} the values of $\kappa$ for the vacuum and thermal cases:
\eqna{
\textrm{Vacuum } ~(T = 0) &:& \quad \kappa = - \eta ~, \IEEEyessubnumber \label{eq:vacuumkappa} \\
\textrm{Thermal } ~(T \neq 0) &:& \quad \kappa = - \eta + z \IEEEyessubnumber \label{eq:thermalkappa} ~.}

As was mentioned in the introduction for the example of wave turbulence, also far from equilibrium there exist important scaling solutions. These solutions can be spacetime translation invariant, however, they are in general not constrained by a fluctuation-dissipation relation. In that case, we may always write down a relation of the form
\eqnlabel{F(p) = -i \left( n(p) + \frac{1}{2} \right) \rho(p) ~,
\label{eq:nFrho}} 
with some generalized `distribution function' $n(p)$ defined from the ratio of $F(p)$ and $\rho(p)$. However, in contrast to \eqref{eq:BE} for the case of thermal equilibrium, here $n(p)$ will in general depend both on frequency $p^{0}$ and spatial momentum $\mbf{p}$. Then from \eqref{ScalingAnsatz1} and \eqref{ScalingAnsatz2} itself no relation between the exponents follows, but only that the distribution function $n(p)$ scales as
\eqn{n(p^{0},\mbf{p}) = s^{\kappa+\eta}\, n(s^{z} p^{0} , s \mbf{p}) ~,}
for $n(p) \gg 1/2$. It will require an actual calculation to determine $\kappa$ at a nonthermal fixed point, and we will show from 
the nonequilibrium renormalization group in a large-$N$ approximation to next-to-leading order that possible scaling solutions are \cite{Berges:2008wm,Berges:2008sr} 
\eqna{\textrm{Nonthermal} : \quad \kappa &=& - \eta + z + d ~ , 
\IEEEyessubnumber \label{eq:particle} \\
\kappa &=& - \eta + 2z + d ~ . \IEEEyessubnumber \label{eq:energy} }
Here, $d$ is the spatial dimension, and we will see that they describe the phenomenon of strong turbulence associated to particle \eqref{eq:particle} and energy \eqref{eq:energy} cascades, respectively \cite{Berges:2008wm,Berges:2010ez,Nowak:2011sk,Berges:2012us}. Because the dimensionality of space enters, the nonthermal values for the exponent $\kappa$ and the corresponding fluctuations can be very large depending on $d$, which has been confirmed also in lattice field theory simulations \cite{Berges:2008wm,Berges:2010ez,Nowak:2010tm,Nowak:2011sk}. Comparing the nonthermal results with \eqref{eq:vacuumkappa} and \eqref{eq:thermalkappa}, one observes a hierarchy of possible fixed point solutions with increasing complexity from vacuum, and thermal equilibrium, to nonequilibrium.

\section{Basics of stationary transport}
\label{sec:weakturbulence}

\subsection{Boltzmann transport}
\vskip\pmsk

Scaling behavior far from equilibrium is typically discussed with the help of kinetic theory or a Boltzmann equation, which can describe transport properties of dilute, weakly interacting 
many-body systems. To make contact with the literature, we discuss the basic concepts of stationary transport in that way before we start from the nonequilibrium renormalization group in quantum field theory in section \ref{sec:NEQFRG}. 

A Boltzmann equation describes the dynamics in terms of a single-particle distribution function $n(t,{\mbf p})$, which depends on time $t$ and spatial momentum ${\mbf p}$ for spatially homogeneous systems. The rate of change in the distribution of particles equals the difference between the rates at which particles in
a phase space region are generated or lost due to collisions. For bosons there is an enhancement of the rate if the final state is already occupied. The collision terms may be expressed
in terms of scattering cross sections and distribution functions. A Boltzmann equation describing $2 \leftrightarrow 2$ scatterings involving four particles reads
\eqna{\partial_{t} n_{\mbf{p}}(t) &=& \int_{\mbf{1},\mbf{2},\mbf{3}}\! \ud\Gamma_{p 1 \leftrightarrow 2 3}  \, \Big[ \big( n_{\mbf{p}} + 1 \big) \left( n_{\mbf{1}} + 1 \right)
n_{\mbf{2}} n_{\mbf{3}} \\ && 
-\: n_{\mbf{p}} n_{\mbf{1}} \left( n_{\mbf{2}} + 1 \right) \left( n_{\mbf{3}} + 1 \right) \Big] \equiv C_{2\leftrightarrow 2}(t,\mbf{p})~,\\ && 
\IEEEyesnumber
\label{Boltzmann}}
where $\int_{\mbf{1},\mbf{2},\mbf{3}}\!\ud\Gamma_{p 1 \leftrightarrow 2 3}$ denotes the relativistically invariant measure to be specified below. We have written $n_{\mbf{p}} = n(t,\mbf{p})$ and $n_{\mbf{i}} = n(t,\mbf{k}_{i})$ as a compact notation for the distribution functions. 

Equation \eqref{Boltzmann} can be understood as arising from the lowest-order perturbative contribution of the respective quantum field theory \cite{Berges:2004yj}. Since the functional renormalization group treatment starting with section \ref{sec:NEQFRG} will include also the perturbative behavior, we only give here some relations to facilitate comparisons with the literature. We consider the example of a relativistic, real scalar field theory with mass $m$ and quartic self-interaction $\lambda$, whose Lagrangian density is 
\eqnlabel{\mathcal{L} = \frac{1}{2} ( \partial_{\mu} \Phi )^{2} - \frac{1}{2} m^{2} \Phi^{2} - \frac{\lambda}{4!} \Phi^{4} ~.
\label{Lagrangian}}
We introduce center and relative (Wigner) coordinates 
\eqn{X = \frac{x + y}{2} ~,  \quad r = x - y ~,}
and Fourier transform both the statistical \eqref{StatisticalFunction} and spectral function \eqref{SpectralFunction} with respect to the relative coordinates
\eqna{
F(X,p) &=& \int\!\ud^{d+1} r\, e^{i p r} F\left(X+r/2, X-r/2\right) ~, \\
\rho(X,p) &=& \int\!\ud^{d+1} r\, e^{i p r} \rho\left(X+r/2, X-r/2\right) ~.}
The Fourier transform of the statistical function is real and that of the spectral function is purely imaginary due to their anti-commutator and commutator definitions, respectively. 

For spatially homogeneous systems, the correlation functions only depend on time $t \equiv X^0$ and four-momentum $p$. We can define a time-dependent `distribution function' $n(t,p)$, depending on four-momentum $p$, by writing 
\eqnlabel{F(t,p) = - i \left( n(t,p) + \frac{1}{2} \right) \rho(t,p)~,
\label{FDRCenterCoordinates}}
which reduces to \eqref{eq:nFrho} for time translation invariant systems. If the spectral function is taken to be of the translation invariant lowest-order (free-field) form 
\eqnlabel{\rho^{(0)}(p) = 2 \pi i \sgn\big(p^{0}\big) \, \delta \big( (p^{0})^{2} - \omega_{\mbf{p}}^{2} \big) ~,
\label{eq:freerho}}
with single-particle energy $\omega_{\mbf p}$, then 
\eqnlabel{n_{\mbf{p}}( t) \equiv - i \int_{0}^{\infty} \frac{\ud p^{0}}{2\pi}\, 2 p^{0} \rho^{(0)}(p)\, n(t,p) ~,
\label{eq:defn}}
corresponds to the distribution function employed in the Boltzmann equation \eqref{Boltzmann}. The definition \eqref{eq:defn} ensures that the single-particle distribution function $n_{\mbf p}(t)$ is evaluated for on-shell four-momentum with only positive energy, i.e.\ $p^0 = \omega_{\mbf p}$. The relativistically invariant measure appearing in \eqref{Boltzmann} is then, for the theory \eqref{Lagrangian}, given by
\eqna{\int_{\mbf{1},\mbf{2},\mbf{3}}\!\ud\Gamma_{p 1 \leftrightarrow 2 3} &=& \frac{\lambda^{2}}{6} \frac{1}{2 \omega_{\mbf{p}}} \,\int \left(\prod_{i = 1}^{3} \frac{\ud^{d} k_{i}}{(2 \pi)^{d}} \frac{1}{2 \omega_{i}} \right)
 \\ && \times\: (2 \pi)^{d+1} \delta^{(d+1)}(p + k_{1} - k_{2} - k_{3}) ~, \\
&& \IEEEyesnumber
\label{BoltzmannMeasure}}
where to lowest order $\omega_{\mbf{p}} = \sqrt{\mbf{p}^2 + m^2}$ and we write $\omega_{\mbf{i}} = \omega(\mbf{k}_{i})$.

More precisely, in a gradient expansion to lowest order in the number of derivatives with respect to the center coordinate $X^0$ and in powers of the relative coordinate $r^0$, the spectral function for spatially homogeneous systems obeys \cite{Berges:2004yj} 
\begin{equation}
p^0 \frac{\partial}{\partial X^0} \rho(X^0,p) = 0 ~ ,
\end{equation}
in agreement with the constant free-field form \eqref{eq:freerho}. 
In contrast, the statistical function $F(t,p)$ to this order can be time-dependent and its evolution is given by 
\eqna{
&&\hspace{-20pt}\int_0^\infty \frac{{\mathrm d} p^0}{2\pi}\, 2 p^0 \partial_{t} F(t,p) \\
&=& -i \int_0^\infty \frac{{\mathrm d} p^0}{2\pi}\, 2  p^0 \rho_p\, \partial_{t} n(t,p) = C(t,\mbf{p}) ~, \IEEEyesnumber \label{eq:genev}}
which defines the collision term $C(t,\mbf{p})$ describing the effects of interactions to lowest order in the gradient expansion. The restriction of the collision term to $2 \leftrightarrow 2$ scatterings, which are the leading perturbative contributions $\sim \lambda^2$, gives the Boltzmann equation \eqref{Boltzmann}. 

We emphasize that the Boltzmann equation has a limited range of applicability. In particular, it is restricted to the perturbative regime. Even for weak coupling, $\lambda \ll 1$, nonperturbative corrections can play a crucial role once the momentum modes become highly occupied. More precisely, equation \eqref{Boltzmann} can only be applied for parametrically small occupancies, $n_{\mbf{p}} \ll 1/\lambda$. Nonperturbatively large occupation numbers $n_{\mbf{p}} \sim 1/\lambda$ lead to important corrections, since the `thin-gas' approximation underlying the Boltzmann equation no longer holds and multiparticle scatterings have to be taken into account. The calculation of these nonperturbative corrections will be a central topic for the nonequilibrium functional renormalization group below. However, for the moment we will consider what one expects from perturbation theory. 

\begin{figure}[t]
\centering
\includegraphics[width=0.49\textwidth]{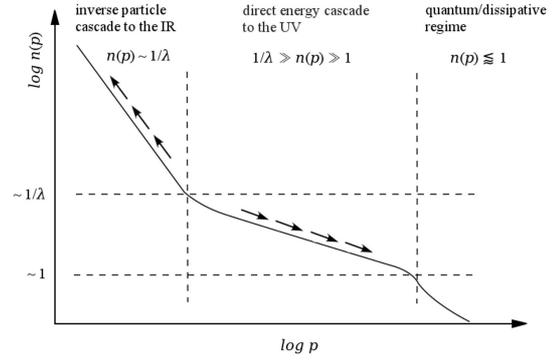}
\caption{Illustration of the dual cascade for scalar quantum field theory. Different scaling exponents can occur in different momentum regimes, which are associated to stationary transport of conserved quantities.}
\label{fig2}
\end{figure}

More precisely, we will analyze \eqref{Boltzmann} in the `classical' regime of occupation numbers \mbox{$1/\lambda \gg n_{\mbf{p}} \gg 1$}, where the collision term can be approximated by
\eqna{C_{2\leftrightarrow 2}^{\rm (cl)}(t,\mbf{p}) &=& \int_{\mbf{1},\mbf{2},\mbf{3}}\! \ud\Gamma_{p 1 \leftrightarrow 2 3}  \, \Big[ ( n_{\mbf{p}} + n_{\mbf{1}} )
n_{\mbf{2}} n_{\mbf{3}} \\ && 
-\: n_{\mbf{p}} n_{\mbf{1}} ( n_{\mbf{2}} + n_{\mbf{3}} ) \Big]~.
\IEEEyesnumber
\label{BoltzmannCl}}
For $n_{\mbf{p}} \lesssim 1$ quantum or `dissipative' processes will play an important role, which will obstruct scaling at sufficiently high momenta. The situation for the real scalar field theory is schematically summarized in Fig.~\ref{fig2}, which sketches the occupation number as a function of momentum on a double-logarithmic scale such that straight lines correspond to power-law behavior. Different slopes can occur in different momentum regions.
The fact that each scaling exponent is associated to an approximately conserved quantity, such as energy or particle number, will be explained next for the perturbative regime using the Boltzmann equation.   

\subsection{Weak wave turbulence}
\vskip\pmsk

We may write the Boltzmann equation \eqref{Boltzmann} formally as a continuity equation
\eqnlabel{\partial_{t} \varepsilon + \nabla_{\mbf{p}} \cdot \mbf{j}_{\mbf{p}} = 0 ~, 
\label{ContinuityEquation}}
for the energy density $\varepsilon(\mbf{p},t) = \omega_{\mbf{p}} n(\mbf{p},t)$ in momentum space, where the divergence of the energy flux $\mbf{j}_{\mbf{p}}$ is given for the collision term \eqref{BoltzmannCl} by
\eqn{\nabla_{\mbf{p}} \cdot \mbf{j}_{\mbf{p}} = - \omega_{\mbf{p}} \,C_{2 \leftrightarrow 2}^{\rm (cl)} (t,\mbf{p})~.}
Since total energy is conserved for the relativistic quantum field theory, we may ask for its effect on the dynamics. We consider the case of an isotropic system where the only nonvanish\-ing part of the flux is given by its radial component. Integrating the continuity equation \eqref{ContinuityEquation} over the volume of the sphere $B(k)$ of radius $k$ in momentum space, we obtain
\eqnlabel{\int_{B(k)} \ud^{d}p \, \partial_{t} \varepsilon = -(2\pi)^{d} A(k) ~,
\label{EnergyFlux}}
which states that the change of net energy contained in $B(k)$ is given by the flux $A(k)$ through the boundary $\partial B(k)$:
\eqnlabel{-(2\pi)^{d} A(k) = \frac{2 \pi^{d/2}}{\Gamma\left(d/2\right)} \int_{0}^{k} \ud|\mbf{p}| \, |\mbf{p}|^{d-1} \omega_{\mbf{p}} \,C_{2 \leftrightarrow 2}^{\rm (cl)} (t,\mbf{p}) ~.
\label{Ak}}
Thermal equilibrium is characterized by the vanishing of the collision term and a zero net flux $A(k)$. Here, however, we are interested in possible stationary solutions of \eqref{Boltzmann} where the distribution $n_{\mbf{p}}$ characterizes some steady state that is far from equilibrium, known as weak wave turbulence \cite{Zakharov:1992}. 
For such a steady state to exist, the distribution function $n_{\mbf{p}}$ needs to satisfy the stationarity condition
\eqnlabel{C_{2\leftrightarrow 2}^{\rm (cl)} [n] = 0~.
\label{FluxState}}
However, in contrast to thermal equilibrium, 
it has a \emph{nonvanishing flux}. Such a `flux state' describes the stationary transport of conserved quantities \cite{Zakharov:1992}. 

Here, we show that these stationary states correspond to the situation where $A(k)$ becomes scale-inde\-pend\-ent. To investigate the behavior of the energy flux under scaling transformations, we make for the single-particle energy a scaling \emph{ansatz} 
\eqn{\omega(s \mbf{p}) = s \omega(\mbf{p}) ~,}
using the linear dispersion for the relativistic scalar field theory \eqref{Lagrangian} at sufficiently high momenta \mbox{$\mbf{p}^2 \gg m^2$}. Similarly, the occupation number distribution is taken to obey the scaling form
\eqn{n(s \mbf{p}) = s^{-\kappa} n(\mbf{p}) ~,}
with the occupation number scaling exponent $\kappa$ introduced already in section \ref{sec:basics}. Together with the scaling property of the measure,
\eqna{&&\hspace{-30pt} \int_{1,2,3} \ud\Gamma_{p 1\leftrightarrow 2 3} (s \mbf{p} , s \mbf{k}_{1} , s \mbf{k}_{2} , s \mbf{k}_{3} ) \\ &=& s^{\mu_4} \int_{1,2,3} \ud\Gamma_{p 1\leftrightarrow 2 3} (\mbf{p} , \mbf{k}_{1} , \mbf{k}_{2}, \mbf{k}_{3}) ~, \IEEEyesnumber
\label{mu4}}
with $\mu_{4} = (3d - 4) - (d+1) = 2 d - 5$ for the scalar field theory case \eqref{BoltzmannMeasure},
we see that the collision integral satisfies
\eqn{C_{2\leftrightarrow 2}^{\rm (cl)} (s \mbf{p}) = s^{-3 \kappa + \mu_{4}}\, C_{2\leftrightarrow 2}^{\rm (cl)} (\mbf{p}) ~.}
This result can easily be generalized for $m$-particle scattering processes
\eqn{C_{m}^{\rm (cl)} (s \mbf{p}) = s^{-\kappa (m-1) + \mu_{m}} C_{m}^{\rm (cl)} (\mbf{p}) ~,}
where $C_{m}^{\rm (cl)}$ denotes the collision term for the case of $m$-particle scattering in the classical regime, and $\mu_{m} = (m-2) d - m - 1$. Using, accordingly, $\omega(\mbf{p}) = |\mbf{p}|\, \omega(1)$ and $n(\mbf{p}) = |\mbf{p}|^{-\kappa}\, n(1)$ etc.\ for the isotropic system, the flux \eqref{Ak} can be seen to give
\eqnlabel{A(k) ~ \sim ~ \frac{\omega(1)\, C_{m}^{\rm (cl)} (1)}{d+1 - \kappa (m-1) + \mu_{m}} \, k^{d + 1 - \kappa (m-1) + \mu_{m}} ~.
\label{eq:fluxexplicit}}
For $A(k)$ to become independent of the scale $k$ up to logarithmic corrections, the $k$-exponent $d + 1 - \kappa (m-1) + \mu_{m}$ must vanish. For the scaling exponent $\kappa$ this gives
\eqnlabel{\kappa = \frac{\mu_{m} + d + 1}{m-1} = d - \frac{m}{m-1}~.
\label{KappaEnergy}}
Since the denominator of \eqref{eq:fluxexplicit} also vanishes in this case, the limit
\eqn{\lim_{\kappa \rightarrow d - m/(m-1)}\, \frac{C_{m}^{\rm (cl)}}{d+ 1 - \kappa (m-1) + \mu_{m}} = const. ~}
has to exist. Therefore, the collision integral must have a zero of first degree in $d+ 1 - \kappa (m-1) + \mu_{m}$ \cite{Zakharov:1992,Berges:2008sr}.
Here, we will assume that the stationary state exists and refer to the literature for further details. Specifically, for the scalar theory with quartic self-interaction in $d = 3$ spatial dimensions, we have the scaling exponent
\eqn{d = 3 ~: \quad \kappa = \frac{5}{3} ~,}
for the transport of energy over some range of momentum scales. This is the so-called energy cascade, which is well-known from the perturbative theory of weak wave turbulence \cite{Zakharov:1992,Micha:2004bv}. We also note that from the relativistic scaling \emph{ansatz} one observes that the scaling exponent associated to momentum conservation is the same as for the energy cascade.

If the dynamics is approximated by the Boltzmann equation \eqref{Boltzmann}, i.e.\ if only the perturbatively leading $2\leftrightarrow 2$ scattering processes are taken into account, then particle number is conserved. Of course, there are total particle number changing processes in the considered relativistic quantum field theory. However, these processes appear at higher order in the coupling, such that for $\lambda \ll 1$ inelastic scattering rates are much smaller than the elastic ones \cite{Jeon:1994if}. Because of this separation of scales, there can be important consequences of approximate particle number conservation for the phenomenon of weak wave turbulence. 
Since the perturbative particle flux is simply given by the momentum integral of the collision integral $C_{2\leftrightarrow 2}^{\rm (cl)}$, going through the corresponding steps as above one may easily verify that the scaling exponent
\eqn{\kappa = d - \frac{m+1}{m-1} ~,}
describes the stationary transport of particles. For the special case of $m = 4$, and $d = 3$, this particle cascade is characterized by the exponent 
\eqn{d = 3 ~: \quad \kappa = \frac{4}{3}~.}

So far, we have only considered effects of $2\rightarrow 2$ particle scattering. For stationary turbulence in quantum field theories, however, the dynamics can lead to a nonvanishing field expectation value or Bose condensation far from equilibrium \cite{Berges:2012us}.
This gives rise to an effective three-vertex ($m=3$) such that the energy cascade in three dimensions is characterized by the scaling exponent~\cite{Micha:2004bv}
\eqn{d = 3 ~: \quad \kappa = \frac{3}{2} ~.}
In these lectures, we will consider only the case of a vanishing field expectation value for simplicity.

From the above discussion we have observed that there are two types of perturbative power-law distributions corresponding to stationary transport of energy and particle number, respectively. These may be realized in different regions of momentum space. However, the simple analysis cannot determine in which regions of momentum space the different scaling solutions are realized, or whether the cascades describe transport from small to large wavenumbers or vice versa. A thorough discussion
leads to the picture of a dual cascade in scalar quantum field theory as illustrated in Fig.~\ref{fig2}~\cite{Berges:2008wm,Berges:2012us,Scheppach:2009wu,Nowak:2010tm,Nowak:2011sk}. Its quantitative description at low momenta requires, however, to go beyond perturbation theory. It has been found that the approximate conservation of an \emph{effective} particle number can be applied to the nonperturbative regime of scalar field theories at low momenta, which leads to different values of turbulent scaling exponents than the perturbative analysis suggests \cite{Berges:2008wm,Berges:2010ez}. Thus, we are in need of reliable nonperturbative techniques to access all characteristic momentum regions of the nonequilibrium dynamics.
In the following, we will consider the functional renormalization group as an ideal tool to access a wide range of scales in a unified framework.

\section{Functional renormalization group}
\label{sec:NEQFRG}

We want to obtain information about scaling solutions for correlation functions far from equilibrium. In the previous section, starting from perturbative kinetic theory or the Boltzmann equation, we illustrated how such a scaling behavior may arise in a weak\-ly coupled scalar theory. However, in the infrared, where occupation numbers become large and the system is strong\-ly correlated, the perturbative description is insufficient. Thus, we need a general framework to calculate correlation functions from first principles. Here, we write down a generating functional for nonequilibrium correlation functions \cite{Calzetta:1986cq,Berges:2004yj}, which serves as a starting point to define the nonequilibrium functional renormalization group on a closed time-path following the presentation of Ref.~\cite{Berges:2008sr}. 

\subsection{Generating functional}
\vskip\pmsk

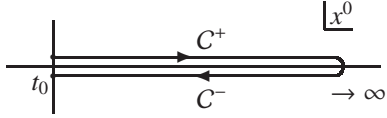
\begin{figure}
\setlength{\unitlength}{0.7pt}
\begin{picture}(20,58) (-45,10)
\put(0,25){\line(1,0){200}}
\put(25,0){\line(0,1){50}}
\put(170,45){\line(0,1){15}}
\put(170,45){\line(1,0){15}}
\thicklines
\put(25,30){\vector(1,0){75}}
\put(100,30){\line(1,0){75}}
\put(175,20){\vector(-1,0){75}}
\put(25,20){\line(1,0){75}}
\put(25,30){\line(1,0){150}}
\put(25,20){\line(1,0){150}}
\cCircle[30](175,25){5}[r]
\put(25,20){\circle*{2.4}}
\put(25,30){\circle*{2.4}}
\put(188,10){\makebox(0,0){$\rightarrow \infty$}}
\put(110,40){\makebox(0,0){$\mathcal{C}^{+}$}}
\put(110,8){\makebox(0,0){$\mathcal{C}^{-}$}}
\put(179,54){\makebox(0,0){$x^{0}$}}
\put(18,15){\makebox(0,0){$t_{0}$}}
\end{picture}
\caption{Closed time path $\mathcal{C}$.}
\label{fig3}
\end{figure}

All information about a nonequilibrium quantum field
theory can be efficiently described in terms of the nonequilibrium generating functional
for correlation functions \cite{Calzetta:1986cq,Berges:2004yj}. For given density matrix $\varrho_0 \equiv \varrho(t_{0})$ at some initial time $t_{0}$, correlation functions can be obtained from the generating functional
\eqna{Z[J, R; \varrho_{0}] &=&  \Tr \varrho_{0} \, \,\textrm{T}_{\mathcal{C}} \exp i \, \left\{ \int_{x,\mathcal{C}} \Phi(x) J(x) \right. \\
&& +\: \left. \frac{1}{2} \int_{x,y,\mathcal{C}} \Phi(x) R(x,y) \Phi(y) \right\} ~, \IEEEyesnumber\IEEEeqnarraynumspace
\label{GeneratingFunctional}}
for a scalar field theory as described by \eqref{Lagrangian} in the presence of sources $J$ and $R$. The introduction of the bilinear source term \mbox{$\sim R$} will be convenient for the derivation of the functional renormalization group equation, which is explained below. Here the time integration is taken over a closed time path $\mathcal{C}$, i.e.\ $\int_{x,\mathcal{C}} \equiv \int_{\mathcal{C}} \ud x^{0} 
\int \ud^{d} x$, displayed in Fig.~\ref{fig3} \cite{Keldysh:1964ud,Schwinger:1960qe}. The closed time-path appears because we want to compute correlation functions, which are given as the \emph{trace} over the density matrix with time-ordered products of Heisenberg field operators, as exemplified in section \ref{sec:basics}. Representing the trace as a path integral will require a time path where the initial and final times are identified, which is discussed in more detail below.  

Above,  $\,\textrm{T}_{\mathcal{C}}$ denotes time-ordering along the contour $\,\mathcal{C}$. As seen from Fig.~\ref{fig3}, this contour consists of an upper $\mathcal{C}^{+}$ and a lower branch $\mathcal{C}^{-}$ where the time-ordering on the lower branch is reversed. To extract correlation functions efficiently, the field $\Phi(x)$ may be written 
in terms of $\Phi^{\pm}(x^{0},\mbf{x})$ where the $\pm$-index denotes on which part of the contour $\mathcal{C}^{\pm}$ the time argument is located. E.g.\ the contour integration for the source term in \eqref{GeneratingFunctional} takes the form
\eqna{&& \hspace{-20pt} \int_{x,\mathcal{C}} \!\Phi(x) J(x) \\
 &\equiv& \int_{x,\mathcal{C}^{+}} \!\! \Phi^{+}(x) J^{+}(x) + \int_{x,\mathcal{C}^{-}} \!\!\Phi^{-}(x) J^{-}(x) \\
&=& \!\int_{t_{0}}^{\infty} \!\!\ud x^{0} \!\int \ud^{d}x \left( \Phi^{+}(x) J^{+}(x) - \Phi^{-}(x) J^{-}(x) \right) ~,}
where the minus sign comes from the reversed time-order\-ing along $\mathcal{C}^{-}$.  
Setting the sources $J$ and $R$ to zero in \eqref{GeneratingFunctional} we obtain the partition sum
\eqn{Z[J,R; \varrho_{0}] \big|_{J,R = 0} = \Tr \varrho_{0} = 1 ~,}
from the normalization of the density matrix.

For instance, taking the second functional derivative of the generating functional \eqref{GeneratingFunctional} with respect to the classical source $J^{+}$ defined on $\mathcal{C}^{+}$
and setting all sources $J$ and $R$ to zero, we obtain
\eqn{\left. \frac{\delta^{2} Z[J , R ; \varrho_{0}]}{i\delta J^{+}(x) \, i \delta J^{+}(y)} \right|_{J,R = 0} = \left\langle \textrm{T}_0 \Phi(x) \Phi(y) \right\rangle \equiv G^{++}(x,y)~.}
Here we have used that $\textrm{T}_{\mathcal{C}}$ is identical to standard time ordering $\textrm{T}_0$ in this case, since both $x^{0}$ and $y^{0}$ lie on the upper part of
the contour, i.e.\ on $\mathcal{C}^+$. We also introduced the notation $G^{++}(x,y)$ in order to distinguish this correlator from the other possible second functional derivatives with respect to the sources $J^{+}$,$J^{-}$ and setting $J,R = 0$ afterwards. These can be written as:
\eqna{G^{++}(x,y) &= & \left\langle \Phi(x) \Phi(y) \, \theta(x^{0} - y^{0}) \right.\\ 
&& \left. +\: \Phi(y) \Phi(x) \, \theta(y^{0} - x^{0}) \right\rangle ~,\\
G^{--} (x,y) &= & \left\langle \Phi(x) \Phi(y) \, \theta(y^{0} - x^{0}) \right. \\
&& \left. +\: \Phi(y) \Phi(x) \, \theta(x^{0} - y^{0}) \right\rangle ~,\\
G^{+-} (x,y) &= & \avr{\Phi(y) \Phi(x)} ~,\\
G^{-+} (x,y) &= & \avr{\Phi(x) \Phi(y)} ~.}
We emphasize that not all of the above two-point functions are independent. In particular, using the property
$\theta(x^{0}-y^{0}) + \theta(y^{0}-x^{0}) = 1$ of the Heaviside step function one obtains
the algebraic identity:
\eqnlabel{G^{++} (x,y) + G^{--}(x,y) = G^{+-}(x,y) + G^{-+}(x,y)~.
\label{AlgebraicIdentity}}
This identity will be of use later on. 

\subsection{Functional integral}
\vskip\pmsk

To simplify the evaluation of correlation functions we write the generating functional \eqref{GeneratingFunctional} in terms of a functional integral representation. For an intuitive presentation we follow standard techniques (see e.g.\ \cite{Chou:1984es}): We evaluate the trace using eigenstates of the Heisenberg field operators $\Phi^{\pm}$ at initial time $t_0$,
\eqn{\Phi^{\pm}(t_{0},\mbf{x}) \, \rbr{\varphi^{\pm}} = \varphi_{0}^{\pm}(\mbf{x}) \,\rbr{\varphi^{\pm}} ~,}
such that \eqref{GeneratingFunctional} may be written as
\eqna{&&\hspace{-15pt} Z[J,R; \varrho_{0}] \\
&=& \int [\ud \varphi_{0}^{+}] \, \lbr{\varphi^{+}} \: \varrho_{0} \, \textrm{T}_{\mathcal{C}} \,\exp i \left\{ \int_{x, \mathcal{C}} \Phi(x) J(x) \right. \\
&& +\: \left. \frac{1}{2} \int_{x,y, \mathcal{C}} \Phi(x) R(x,y) \Phi(y) \right\} \rbr{\varphi^{+}} ~. \IEEEyesnumber\IEEEeqnarraynumspace
\label{GeneratingFunctional1}}
Here the integration measure is given by
\eqnlabel{\int [\ud\varphi_{0}^{\pm}] \, \equiv \int \prod_{\mbf{x}} \ud\varphi_{0}^{\pm}(\mbf{x}) ~.
\label{FunctionalMeasure}}
With the insertion 
\eqnlabel{\int [\ud \varphi_{0}^{-}] \:\rbr{\varphi^{-}} \lbr{\varphi^{-}} = \mathds{1}~,
\label{UnitOperator}}
we may bring \eqref{GeneratingFunctional1} to a form 
\eqna{Z[J,R; \varrho_{0}] &=& \int [\ud \varphi_{0}^{+}] [\ud \varphi_{0}^{-}] \: \lbr{\varphi^{+}} \, \varrho_{0} \, \rbr{\varphi^{-}} \\
&& \times\: \left( \varphi^{-},t_{0} \,|\, \varphi^{+},t_{0} \right)_{J,R} ~.  \IEEEyesnumber\IEEEeqnarraynumspace
\label{GeneratingFunctional2}}
Here the transition amplitude in the presence of the sources is given by
\eqna{&& \hspace{-25pt}\left( \varphi^{-},t_{0} \,|\, \varphi^{+},t_{0} \right)_{J,R} \\ 
&\equiv& \lbr{\varphi^{-}} \, \textrm{T}_{\mathcal{C}} \,\exp i \left\{ \int_{x, \mathcal{C}} \Phi(x) J(x) \right. \\
&& +\: \left. \frac{1}{2} \int_{x,y, \mathcal{C}} \Phi(x) R(x,y) \Phi(y) \right\} \, \rbr{\varphi^{+}} ~. 
\IEEEyesnumber\IEEEeqnarraynumspace
\label{MatrixElement}} 
This matrix element can be written as a functional integral over the fields $\varphi^{\pm}$
\eqna{&& \hspace{-18pt} \left( \varphi^{-},t_{0} \,|\, \varphi^{+},t_{0} \right)_{J,R} \\ &=& \int [\ud\varphi^{+}]' [\ud\varphi^{-}]'  \exp i \left\{ S[\varphi] + \int_{x,\mathcal{C}} \varphi(x) J(x) \right.\\
&& +\: \left. \frac{1}{2} \int_{x,y,\mathcal{C}} \varphi(x) R(x,y) \varphi(y) \right\}~,
\IEEEyesnumber\IEEEeqnarraynumspace
\label{GeneratingFunctional2b}}
which is essentially the same procedure as employed to obtain standard path integral expressions for vacuum of equilibrium matrix elements \cite{Landsman:1986uw}. Here $S[\varphi]$ is the classical action for the scalar theory, and the measure is given by
\eqnlabel{\int [\ud\varphi^{\pm}]' \, \equiv \int\!\prod_{\mbf{x}; \, x^{0} \,>\, t_{0}}\! \ud\varphi^{\pm} (x^{0},\mbf{x}) ~.
\label{FunctionalMeasure2}}
Here, the functional integration goes over the field configurations $\varphi^{\pm}(x^{0},\mbf{x})$ that satisfy the boundary condition $\varphi^{\pm}(x^{0} = t_{0},\mbf{x}) = \varphi_{0}^{\pm}(\mbf{x})$.

The above expression \eqref{GeneratingFunctional2} together with \eqref{GeneratingFunctional2b} displays two important ingredients entering nonequilibrium quantum field theory: the quantum fluctuations described by the functional integral with action $S$ that determines the transition matrix element, and the statistical fluctuations encoded in the averaging procedure over the initial conditions as specified by the initial density matrix~$\varrho_{0}$.

\subsection{$N$-component scalar field theory}
\vskip\pmsk

So far, we have not specified any internal field degrees of freedom. In the following, we consider a $N$-component vector field $\Phi_{a}$ with $a = 1, \ldots , N$ for an $O(N)$-symmetric theory. The number of field components will later serve as an expansion parameter that allows for a controlled approximation of renormalization group equations. The classical action for the $O(N)$-model with a quartic self-interaction is given by
\eqna{S[\varphi] &=& \frac{1}{2} \int_{x,y,\mathcal{C}} \varphi_{a} (x) i D^{-1}_{a b} (x,y) \varphi_{b} (y) \\ && -\: \frac{\lambda}{4 N!} \int_{x,\mathcal{C}} \varphi_{a} (x) \varphi_{a} (x) \varphi_{b} (x) \varphi_{b} (x) ~, \\ && \IEEEyesnumber
\label{ActionSchwingerKeldysh}}
where the time integration runs over the contour and $i D_{a b}^{-1}$ denotes the free inverse propagator satisfying
\eqn{\left( - \Box_{x} - m^{2} \right) i D_{a b} (x,y) = \delta_{a b} \delta_{\mathcal{C}}^{(d+1)}(x-y) ~.}
Using again the $\pm$-index to denote on which part of the contour $\mathcal{C}^{\pm}$ the time argument is located, the free part of the action takes the form
\eqna{ S_{0}[\varphi^{+},\varphi^{-}]  &=& \frac{1}{2} \int_{x,y} \left( \varphi_{a}^{+} (x) i D^{-1}_{a b} (x,y)  \varphi_{b}^{+} (y) \right. \\
&& 
\left. -\: \varphi_{a}^{-} (x) i D^{-1}_{a b} (x,y) \varphi_{b}^{-}(y) \right) ~.}
The interaction term is written as
\eqna{S_{\rm int} [\varphi^{+},\varphi^{-}] &=& - \frac{\lambda}{4 N!} \int_{x} \left( \varphi_{a}^{+}(x) \varphi_{a}^{+}(x) \varphi_{b}^{+}(x) \varphi_{b}^{+}(x) \right. \\
&& 
\left. -\: \varphi_{a}^{-}(x) \varphi_{a}^{-}(x) \varphi_{b}^{-}(x) \varphi_{b}^{-}(x) \right)~.}
The time integration is implicitly defined along the positive branch of the contour, i.e.\ $\int_{x} \, \equiv \int_{t_{0}}^{\infty} \ud x^{0} \int \ud^{d} x$ and the minus sign in front of the $\varphi^{-}$-components of the action comes from the reversed time ordering along the lower branch $\mathcal{C}^{-}$.

\subsection{Quantum vs.\ classical dynamics}
\vskip\pmsk

In order to discuss the different origins of quantum and of classical-statistical fluctuations, it is convenient to introduce\footnote{In order to prevent a proliferation of symbols, the notation does not distinguish the new definition of $\varphi_a$ from its earlier use since no confusion for the following can occur.}
\eqnlabel{\varphi_{a} = \frac{1}{2} \left( \varphi_{a}^{+} + \varphi^{-}_{a} \right) ~, \quad {\tilde \varphi}_{a} = \varphi_{a}^{+} - \varphi_{a}^{-} ~.
\label{Basis}}
For the rest of the lectures we will use this basis for our calculations. The action \eqref{ActionSchwingerKeldysh} takes the following form
\eqnlabel{S[\varphi,{\tilde \varphi}] = S_{0} [\varphi , {\tilde \varphi}] + S_{\rm int} [\varphi,{\tilde \varphi}] ~,
\label{ActionRetardedAdvanced}}
where the free part is given by
\eqnlabel{S_{0} [\varphi,{\tilde \varphi}] = \int_{x,y} {\tilde \varphi}_{a} (x)  i D_{a b}^{-1} (x,y) \varphi_{b} (y) ~,
\label{FreePartAction}}
and the interaction part reads
\eqna{S_{\rm int}[\varphi,{\tilde \varphi}] &=& - \frac{\lambda}{6 N} \int_{x} {\tilde \varphi}_{a} (x) \varphi_{a} (x) \varphi_{b} (x) \varphi_{b} (x) 
\\
&& - \frac{\lambda}{24 N} \int_{x} {\tilde \varphi}_{a} (x) {\tilde \varphi}_{a} (x) {\tilde \varphi}_{b} (x) \varphi_{b} (x) ~,  \IEEEyesnumber\IEEEeqnarraynumspace
\label{InteractionPartAction}}

Finally, the linear source term can be written in this basis takes the form
\eqna{&& \hspace{-30pt}\int_{x} \!\left( \varphi^{+}_{a}(x) J^{+}_{a} (x) - \varphi^{-}_{a} (x) J^{-}_{a} (x) \right) \\
&=& \int_{x} \left( \varphi_{a}(x) {\tilde J}_{a} (x) + {\tilde \varphi}_{a} (x) J_{a} (x)  \right)~,}
whereas the bilinear source term is written as
\eqna{&& \hspace{-15pt} \int_{x,y} \left( \varphi^{+}_{a} (x) \,,\, \varphi^{-}_{a} (x) \right) 
\begin{pmatrix} R^{++}_{a b} (x,y) & - R^{+-}_{a b} (x,y) \\[\pmsk] - R^{-+}_{a b} (x,y) & R^{--}_{a b} (x,y)\end{pmatrix} 
\begin{pmatrix} \varphi^{+}_{b} (y) \\[\pmsk] \varphi_{b}^{-} (y) \end{pmatrix} \\
\hspace{0pt} &=& \int_{x,y} \!\left( \varphi_{a} (x) \,,\, {\tilde \varphi}_{a} (x) \right) 
\begin{pmatrix} R^{\tilde F}_{a b} (x,y) & R^{\rmA}_{a b} (x,y) \\[\pmsk] R^{\rmR}_{a b} (x,y) & R^{F}_{a b} (x,y)\end{pmatrix} 
\begin{pmatrix} \varphi_{b} (y) \\[\pmsk] {\tilde \varphi}_{b} (y) \end{pmatrix} ~.}
We have added the indices $\rmR,\rmA,F,{\tilde F}$ to the bilinear sources that suggest a relation to the retarded, advanced, and statistical components in the new basis. This connection will be made more explicit in the following subsections.

\begin{figure}[!t]
\centering
\begin{picture}(20,42) (295,-164)
    \SetWidth{0.5}
    \SetColor{Black}
\Text(242,-135)[lb]{$\varphi $}
\Text(273,-135)[lb]{$\varphi $}
\Text(242,-164)[lb]{${\tilde \varphi}$}
\Text(273,-164)[lb]{$\varphi $}
    \Text(333,-135)[lb]{${\tilde \varphi}$}
    \Text(363,-135)[lb]{${\tilde \varphi}$}
    \Text(333,-164)[lb]{$\varphi $}
    \Text(363,-164)[lb]{${\tilde \varphi}$}
\SetWidth{1.0}
    \Line(250,-135)(270,-155)
    \Line(260,-145)(270,-135)
    \DashLine(260,-145)(250,-155){2}
\SetWidth{1.0}
    \DashLine(340,-135)(350,-145){2}
    \DashLine(350,-146)(360,-135){2}
    \Line(350,-145)(340,-155)
    \DashLine(360,-155)(350,-145){2}
\end{picture}
\caption{Classical (left) and quantum vertex (right) in the scalar field theory.}
\label{fig4}
\end{figure}
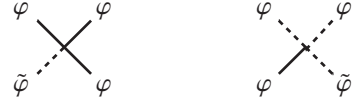

The two types of vertices appearing in the interaction part \eqref{InteractionPartAction} are illustrated in Fig.~\ref{fig4}. To understand their role for the dynamics, it is important to note that one can also write down a functional integral for the corresponding nonequilibrium \emph{classical-statistical} field theory. The standard derivation of the latter employs that the classical field equation of motion can be obtained from $S[\varphi,{\tilde \varphi}]$ by functional differentiation with respect to~$\tilde{\varphi}$,
\eqna{\left. \hspace{-15pt} \frac{\delta S[\varphi,{\tilde \varphi}]}{\delta {\tilde \varphi}_{a}(x)} \right|_{{\tilde \varphi} = 0} &=& -\left( \Box_{x} + m^{2} \right) \varphi_{a} (x) \\
&& -\: \frac{\lambda}{3 N} \varphi_{a} (x) \varphi_{b}(x) \varphi_{b}(x) = 0~, \IEEEyesnumber
\label{ClassicalDynamics}}
where the derivative had to be evaluated for \mbox{$\tilde{\varphi}=0$} to eliminate terms originating from the part of the interaction term \eqref{InteractionPartAction} that is cubic in $\tilde{\varphi}$.
This classical dynamics can then be implemented as a functional integral using the representation of the $\delta$-functional 
\eqna{ && \hspace{-30pt}
\delta\left[ \frac{\delta S[\varphi,{\tilde \varphi}]}{\delta {\tilde \varphi}_{a}(x)} \Big|_{{\tilde \varphi} = 0} \right] \\ 
&=& \int [d \tilde{\varphi}] \exp\left( i \int_x \frac{\delta S[\varphi,{\tilde \varphi}]}{\delta {\tilde \varphi}_{a}(x)} \Big|_{{\tilde \varphi} = 0}{\tilde \varphi}_{a}(x) \right) ~, \IEEEyesnumber
\label{eq:classicalc}}
with the help of the `auxiliary' field $\tilde{\varphi}$ and further steps involving the integration with respect to $\varphi$ \cite{Martin:1973zz,Jeon:2004dh,Berges:2007ym}. We emphasize that in the exponent on the r.h.s.\ of \eqref{eq:classicalc} only terms linear in $\tilde{\varphi}$ appear. In particular, the  interaction term $\sim \tilde{\varphi}^3$, appearing with \eqref{InteractionPartAction} in the functional integral of the quantum theory, does not occur for the classical theory. In summary, the generating functionals for correlation
functions are very similar in the quantum and the classical
statistical theory. A crucial difference is that the quantum
theory is characterized by an additional vertex. In a diagrammatic language, if loop corrections involving only the classical vertex dominate over those involving the quantum vertex or a combination of both, then the quantum dynamics can be approximately described by the classical-statistical field theory. This has been analyzed in great detail in recent years for nonequilibrium phenomena \cite{Aarts:2001yn,Arrizabalaga:2004iw,Berges:2008wm}, and we will come back to this point in the context of turbulence below.

\subsection{Connected one- and two-point functions}
\vskip\pmsk

The generating functional for connected correlation functions is given by
\eqnlabel{W  = - i \ln Z  ~.
\label{GeneratingFunctional3}}
We may calculate field expectation values by functional differentiation with respect to the sources $J$ and $\tilde{J}$, which we denote as
\eqn{\frac{\delta W}{\delta J_{a} (x)} = {\tilde \phi}_{a} (x) ~, 
\quad \frac{\delta W}{\delta \tilde{J}_{a} (x)} = \phi_{a} (x) ~,}
for fixed $J$ and $R$. 
The connected two-point correlation functions are given by the second functional derivatives
\eqna{\frac{\delta^{2} W}{\delta \tilde{J}_{a} (x) \delta J_{b} (y)} &=& G^{\rmR}_{a b} (x,y) ~, 
\\
\frac{\delta^{2} W}{\delta J_{a} (x) \delta \tilde{J}_{b} (y)} &=& G^{\rmA}_{a b} (x,y) ~, 
\\
\frac{\delta^{2} W}{\delta \tilde{J}_{a} (x) \delta \tilde{J}_{b} (y)} &=& i F_{a b} (x,y) ~, 
\\
\frac{\delta^{2} W}{\delta J_{a} (x) \delta J_{b} (y)} &=& i \tilde F_{a b} (x,y)~,
\IEEEyesnumber\IEEEeqnarraynumspace
\label{eq:connectedtwopoint}}
where $G^{\rmR,\rmA}$ are the retarded/advanced propagators, $F$ is the statistical propagator, and ${\tilde F}$ is the `anomalous' propagator. We note that the retarded and advanced propagators satisfy the symmetry property
\eqnlabel{G_{a b}^{\rmA} (x,y) = G^{\rmR}_{b a} (y,x) ~,
\label{eq:arsym}}
and for the statistical propagators, we have
\eqn{F_{a b} (x,y) = F_{b a} (y,x) ~, \quad {\tilde F}_{a b} (x,y) = {\tilde F}_{b a} (y,x) ~.}
These properties follow directly from the definition of the propagators in terms of the second functional derivatives with respect to $J$ and ${\tilde J}$.
The spectral function $\rho$ is given by the difference of the retarded and advanced propagators
\eqn{\rho_{a b} (x,y) = G^{\rmR}_{a b} (x,y) - G_{a b}^{\rmA} (x,y) ~,}
and $G^{\rmR}_{a b} (x,y) = \rho_{a b} (x,y) \theta (x^{0}-y^{0})$.
It is important to note that the anomalous propagator ${\tilde F}$ vanishes in the limit where the external sources are set to zero. This is a consequence of the algebraic identity \eqref{AlgebraicIdentity}, since ${\tilde F}= G^{++} + G^{--} - G^{-+} - G^{+-}$, as one may readily check by changing basis. More generally, in the absence of sources, we have \cite{Chou:1984es}
\eqna{ \frac{\delta W}{\delta J_{a} (x)} \Big|_{J,{\tilde J} , R^{R,A,F,{\tilde F}} = 0} &=& {\tilde \phi}_{a} (x) = 0 ~,
\\
 \frac{\delta^{2} W}{\delta J_{a} (x) \,\delta J_{b} (y)} \Big|_{J,{\tilde J} ,R^{R,A,F,{\tilde F}} = 0} &=& i {\tilde F}_{a b} (x,y) = 0 ~,}
and, correspondingly, arbitrary functional derivatives of the generating functional with respect to $J$ vanish in the absence of sources.

\subsection{Functional renormalization group}
\vskip\pmsk

A most convenient derivation of the functional renormalization group equation starts from the two-particle irreducible (2PI) effective action \cite{Cornwall:1974vz}.
The latter is obtained as a Legendre transform of the generating functional \eqref{GeneratingFunctional3} with respect to the linear and bilinear source terms $J,R$. One obtains a functional of the field expectation values $\phi,{\tilde \phi}$, and the propagators $G^{\rmR},G^{\rmA},F,{\tilde F}$:
\eqna{&& \hspace{-17pt} \Gamma_{2\mathrm{PI}}\left[\phi,{\tilde \phi},G^{\rmR},G^{\rmA},F,{\tilde F}\right] \\
&=& W - \int_{x} \left\{ \phi_{a} (x) \tilde{J}_{a}(x) + {\tilde \phi}_{a} (x) J_{a} (x) \right\} \\
&& \hspace{0pt} -\: \frac{1}{2} \int_{x,y} \bigg\{ R^{\rmA}_{a b} (x,y) \left[ \phi_{a} (x) {\tilde \phi}_{b} (y) - i G^{\rmR}_{a b} (x,y) \right] \\
&& \hspace{31pt} +\: R^{\rmR}_{a b} (x,y) \left[ {\tilde \phi}_{a} (x) \phi_{b} (y) - i G^{\rmA}_{a b} (x,y) \right] \\
&& \hspace{31pt} +\: R^{\tilde F}_{a b} (x,y) \, \Big[ \phi_{a} (x) \phi_{b} (y) + F_{a b} (x,y) \Big] \\
&& \hspace{31pt} +\: R^F_{a b} (x,y) \left[ {\tilde \phi}_{a} (x) {\tilde \phi}_{b} (y) + \tilde F_{a b} (x,y) \right] \bigg\} ~,}
where the sources depend on the field expectation values and the propagators, i.e.\ $J = J\left[\phi,{\tilde \phi}, G^{\rmR},G^{\rmA},F,{\tilde F}\right]$ etc.

We may partially undo the Legendre transform for the bilinear sources $R^{\rmR,\rmA,F,{\tilde F}}$ to obtain an expression for the one-particle irreducible (1PI) effective action in the presence of these sources, i.e.\
\eqna{\hspace{-4pt}\Gamma\left[\phi,{\tilde \phi},R^{\rmR,\rmA,F,{\tilde F}}\right] &=& \Gamma_{2\mathrm{PI}} - \frac{i}{2} \Tr \left\{ G^{\rmR} R^{\rmR} + G^{\rmA} R^{\rmA} \right. \\
&& \left. +\: i F R^{\tilde F} + i \tilde F R^F \right\} ~. \IEEEyesnumber \label{EffectiveAction}}
Here the propagators depend on the fields and bilinear sources, i.e.\ $G^{\rmR}=G^{\rmR}\left[\phi, {\tilde \phi},R^{\rmR,\rmA,F,\tilde F}\right]$ etc.
The above equation is our starting point for the construction of the functional renormalization group. We take the bilinear sources to depend on some characteristic momentum scale $k \geq 0$, which renders the effective action scale-dependent,
\eqn{
\Gamma_{k} [\phi, {\tilde \phi}] \equiv \Gamma\left[ \phi , {\tilde \phi} , R^{\rmR,\rmA,F,\tilde F}_{k} \right].}  
Taking now the derivative with respect to the scale $k$, denoting
\eqn{\dot \Gamma_{k} [\phi, {\tilde \phi}] \equiv k \frac{\partial \Gamma_{k} [\phi, {\tilde \phi}]}{\partial k} ~,} 
we get from \eqref{EffectiveAction} the renormalization group equation for the scale-dependent effective action:
\eqnlabel{\hspace{-2pt}{\dot \Gamma}_{k} = - \frac{i}{2} \Tr \left\{ G_{k}^{\rmR} {\dot R}_{k}^{\rmR} + G_{k}^{\rmA} {\dot R}_{k}^{\rmA} + i F_{k} {\dot R}_{k}^{\tilde F} 
+ i \tilde F_{k} {\dot R}_{k}^F \right\}~,
\label{eq:RGequation}}
Here we have used the fact that $\Gamma_{2\mathrm{PI}}$ in \eqref{EffectiveAction} is independent of the bilinear sources $R_{k}^{\rmR,\rmA,F,\tilde F}$. The flow equation \eqref{eq:RGequation} corresponds to the Wetterich equation for the effective average action \cite{Wetterich:1992yh}, however, now evaluated on the closed time path.

\subsection{Cutoff functions}
\vskip\pmsk

With suitable cutoff functions, the effective average action $\Gamma_k$ may be viewed as a coarse grained effective action, which includes all quantum-statistical fluctuations with characteristic momenta above the scale $k$. If the infrared cutoff scale $k$ is sent to zero, one recovers the standard 1PI effective action with all fluctuations included, i.e. 
\eqnlabel{ \Gamma_{k=0} [\phi,{\tilde \phi}] = \Gamma \left[\phi, {\tilde \phi},R_{k=0}^{\rmR,\rmA,F,{\tilde F}}=0\right] ~.
\label{GammaZero}}
Therefore, the sources $R_{k}^{\rmR,\rmA,F,{\tilde F}}$ must vanish in the infrared limit. 
With this property, the renormalization group equation may be used to describe the flow starting from some high momentum scale $\Lambda$, where the microscopic physics is characterized by some classical action $S$, i.e.
\eqnlabel{\lim_{k \rightarrow \Lambda} \Gamma_{k} [\phi,{\tilde \phi}] \simeq S [\phi, {\tilde \phi}] ~.
\label{GammaLambda}}
The renormalization group flow then interpolates between the classical action and the full quantum effective action that appears when all fluctuations have been taken into account.

We have seen that the renormalization group on the closed time path requires the specification of the various cutoff functions $R_k^{\rmR,\rmA,F,{\tilde F}}$, which seems to allow for a wider class of possible choices than in the corresponding Euclidean field theories. In order to discuss what constraints can be obtained from the requirement \eqref{GammaLambda}, it is useful to start from the -- now $k$-dependent -- generating functional \eqref{GeneratingFunctional}:
\eqna{\hspace{-18pt} Z_{k} [J,{\tilde J}] &=& \int [\ud \varphi] [\ud {\tilde \varphi}] \exp i \bigg\{ S [\varphi, {\tilde \varphi}] \\
&& +\: \int_{x} \left\{ \varphi_{a} (x) \tilde{J}_{a} (x) + {\tilde \varphi}_{a} (x) J_{a} (x) \right\} \\
&& +\: \frac{1}{2} \int_{x,y} \Big\{ \varphi_{a} (x) R^{\rmA}_{k,a b} (x,y) {\tilde \varphi}_{b} (y)  \\
&& \hspace{30pt} +\: {\tilde \varphi}_{a} (x) R^{\rmR}_{k, a b} (x,y) \varphi_{b} (y) \\
&& \hspace{30pt} +\: \varphi_{a} (x) R^{\tilde F}_{k, a b} (x,y) \varphi_{b} (y) \\
&& \hspace{30pt} +\: {\tilde \varphi}_{a} (x) R^F_{k, a b} (x,y) {\tilde \varphi}_{b} (y) \Big\} \bigg\}~.  \IEEEyesnumber
\label{GeneratingFunctional4}}
Here we have hidden the averaging over the initial density matrix $\varrho_{0}$ in the notation, since we will not be concerned with the special choice of the initial conditions for the present purposes. Later on we will comment on the role of the initial conditions. The affective average action \eqref{EffectiveAction} can then be written as
\eqna{\Gamma_{k} [\phi,{\tilde \phi}] &=& - i \ln Z_{k} - \int_{x} \left\{ \phi_{a} (x) \tilde{J}_{a} (x) + {\tilde \phi}_{a} (x) J_{a} (x) \right\} \\
&& -\: \frac{1}{2} \int_{x,y} \bigg\{ \phi_{a} (x) R^{\rmA}_{k,a b} (x,y) {\tilde \phi}_{b} (y)  \\
&& \hspace{30pt} +\: {\tilde \phi}_{a} (x) R^{\rmR}_{k, a b} (x,y) \phi_{b} (y) \\
&& \hspace{30pt} +\: \phi_{a} (x) R^{\tilde F}_{k, a b} (x,y) \phi_{b} (y) \\
&& \hspace{30pt} +\: {\tilde \phi}_{a} (x) R^F_{k, a b} (x,y) {\tilde \phi}_{b} (y) \bigg\}~.
\label{1PIEffectiveAction} \IEEEyesnumber}
In the following, we employ the notation
\eqn{\Gamma_{k,a}^{\phi} (x) \equiv \frac{\delta \Gamma_{k}}{\delta \phi_{a} (x)} ~, \quad   \Gamma^{{\tilde \phi}}_{k, a} (x) \equiv \frac{\delta \Gamma_{k}}{\delta {\tilde \phi}_{a} (x)} ~.}
By functional differentiation of \eqref{1PIEffectiveAction} with respect to the fields $\phi_{a}$ and $\tilde{\phi}_{a}$, we obtain the equations of motion for the fields $\phi$ and $\tilde{\phi}$:
\eqna{\Gamma_{k,a}^{\phi} (x) &=& - \tilde{J}_{a} (x) - \int_{y} \left\{ R^{\tilde F}_{k, a b} (x,y) \phi_{b} (y) \right. \\
&+& \left. \frac{1}{2} R^{\rmA}_{k, a b} (x,y) {\tilde \phi}_{b} (y) + \frac{1}{2} {\tilde \phi}_{b} (y) R^{\rmR}_{k, b a} (y, x) \right\},\\
&& \IEEEyessubnumber\IEEEeqnarraynumspace 
\label{EoM1} \\
\Gamma^{{\tilde \phi}}_{k, a} (x) &=& - J_{a} (x) - \int_{y} \left\{ R^F_{k, a b} (x,y) {\tilde \phi}_{b} (y) \right. \\
&+& \left. \frac{1}{2} R^{\rmR}_{k, a b} (x,y) \phi_{b} (y) + \frac{1}{2} \phi_{b} (y) R^{\rmA}_{k, b a} (y,x) \right\} .\\
&& \IEEEyessubnumber\IEEEeqnarraynumspace
\label{EoM2}}
With these we may eliminate the sources $\tilde{J}_{a} (x)$ and $J_{a} (x)$ from \eqref{1PIEffectiveAction}, expressing them in terms of the first functional
derivatives of $\Gamma_{k}$. We then shift the fields in the generating functional \eqref{GeneratingFunctional4} by
\eqn{\varphi \rightarrow \phi + \varphi ~, \quad {\tilde \varphi} \rightarrow {\tilde \phi} + {\tilde \varphi}~,}
observing that the measure is invariant under these transformations. 
That way, from \eqref{1PIEffectiveAction}, we finally arrive at the following functional integro-differential equation for the effective action:
\eqna{&&\hspace{-12pt} \Gamma_{k}[\phi,{\tilde \phi}] \\ &=& S[\phi,{\tilde{\phi}}] - i \ln \int [\ud\varphi] [\ud{\tilde \varphi}]
\exp i \, \bigg\{ S[\phi + \varphi , {\tilde \phi} + {\tilde \varphi}] \\
&&  \hspace{15pt} -\: S [\phi, {\tilde \phi}] + \int_{x} \left\{ \varphi_{a} (x) \Gamma^{\phi}_{k,a} (x) + {\tilde \varphi}_{a} (x) \Gamma^{{\tilde \phi}}_{k,a} (x) \right\} \\
&& \hspace{15pt} -\: \frac{1}{2} \int_{x,y} \left\{ \varphi_{a} (x) R^{\rmA}_{k,a b} (x,y) {\tilde \varphi}_{b} (y) \right. \\
&& \hspace{44pt} +\: {\tilde \varphi}_{a} (x) R^{\rmR}_{k, a b} (x,y) \varphi_{b} (y) \\
&& \hspace{44pt} +\: \varphi_{a} (x) R^{\tilde F}_{k, a b} (x,y) \varphi_{b} (y) \\
&& \left. \hspace{43pt} +\: {\tilde \varphi}_{a} (x) R^F_{k, a b} (x,y) {\tilde \varphi}_{b} (y) \right\} \bigg\}~.  \IEEEyesnumber
\label{1PIEffectiveAction1}}
With the following representation for the $\delta$-functional
\eqn{\delta[\varphi] = \int [\ud {\tilde \varphi}] \exp \left\{ i \int_{x} \varphi_{a} (x) {\tilde \varphi}_{a} (x) \right\}~,} 
and equivalently for $\delta[{\tilde{\varphi}}]$, one observes that the property \eqref{GammaLambda} can be efficiently achieved with a class of cutoff functions chosen as
\eqna{R^{\rmR,\rmA}_{k, a b} (x,y) &=& R_{k} \left(-\Box_{x}\right) \delta(x-y) \delta_{a b} ~, \\ R^{F,\tilde F}_{k, a b}(x,y) &=& 0~, \IEEEyesnumber
\label{eq:cutoffclass}}
with the property
\eqnlabel{\lim_{k \rightarrow \Lambda} R_{k} \rightarrow \infty ~.
\label{RkInfinity}}
This ensures that the bilinear source terms in \eqref{1PIEffectiveAction1} act as $\delta$-constraints in the limit $k \rightarrow \Lambda$, thus suppressing fluctuations in the fields $\varphi$ and $\tilde{\varphi}$.
More generally, possible nonzero $R_{k}^{F,\tilde{F}}$ should not be chosen to grow as fast as $R^{\rmR,\rmA}_{k}$ for $k \rightarrow \Lambda$ in order to comply with \eqref{GammaLambda} \cite{Berges:2008sr}. 

The choice of vanishing $R_{k}^{F,{\tilde{F}}}$ greatly simplifies the structure of the flow equations.
In this case the exact flow equation for the effective average action \eqref{eq:RGequation} becomes
\eqnlabel{{\dot \Gamma}_{k} [\phi,{\tilde \phi}] = - \frac{i}{2} \Tr \left\{ G^{\rmR}_{k} {\dot R}^{\rmR}_{k} 
+ G^{\rmA}_{k} {\dot R}^{\rmA}_{k} \right\} ~. 
\label{FlowEquation}}
Furthermore, from the symmetry properties \eqref{eq:arsym} of the propagators $G_{k}^{\rmR,\rmA}$, and the cutoff functions $R_{k}^{\rmR,\rmA}$, one observes that $\Tr \left\{ G^{\rmR}_{k} [\phi,{\tilde \phi}] {\dot R}^{\rmR}_{k} \right\} = \Tr \left\{ G^{\rmA}_{k} [\phi,{\tilde \phi}] {\dot R}^{\rmA}_{k} \right\}$ holds. However, it proves to be convenient to keep both the retarded and advanced functions in \eqref{FlowEquation} as it simplifies the diagrammatic rules that will be introduced below.

\subsection{Propagators}
\vskip\pmsk

Above we have derived the exact flow equation for the effective average action $\Gamma_{k}$ that depends on the retarded and advanced propagators $G_{k}^{\rmR}$ and $G_{k}^{\rmA}$. It remains to relate the two-point functions to functional derivatives of $\Gamma_{k}$. Here, we want to illustrate how to obtain these relations starting from simple identities for the sources $J$ and ${\tilde J}$.
As an example, we may consider the following identity
\eqna{&& \hspace{-10pt} \delta^{(d+1)}(x-y) \delta_{a b} \\
&=& \frac{\delta J_{a} (x)}{\delta J_{b} (y)} = \int_{z} \left\{ \frac{\delta J_{a} (x)}{\delta \phi_{c} (z)} \frac{\delta \phi_{c} (z)}{\delta J_{b} (y)} 
+ \frac{\delta J_{a} (x)}{\delta {\tilde \phi}_{c} (z)}\frac{{\tilde \phi}_{c} (z)}{J_{b} (y)} \right\} ~, \\
&& \IEEEyesnumber\IEEEeqnarraynumspace
\label{Rel1}}
where we have used that $J = J[\phi,{\tilde \phi}]$ is a functional of the field expectation values. Using the equation of motion
\eqn{\frac{\delta \Gamma_{k}}{\delta {\tilde \phi}_{a} (x)} = - J_{a} (x) - \int_{y} R^{\rmR}_{k, a b} (x,y) \phi_{b} (y) ~, }
we may write the functional derivatives $\delta J_{a} (x)/\delta \phi_{c} (z)$ and $\delta J_{a} (x)/\delta {\tilde \phi}_{c} (z)$ in terms of second functional derivatives of the effective average action. Here we have exploited the symmetry property of the cutoff functions $R_{k}^{\rmR,\rmA}$ and the vanishing of $R_{k}^{F,{\tilde F}}$ to write \eqref{EoM2} in a somewhat simpler form. Furthermore, the functional derivatives of the field expectation values
\eqn{\frac{\delta \phi_{c} (z)}{\delta J_{b} (y)} = \frac{\delta^{2} W_{k}}{\delta \tilde{J}_{c} (z)\,\delta J_{b} (y)} ~, \quad
\frac{\delta {\tilde \phi}_{c} (z)}{\delta J_{b} (y)} = \frac{\delta^{2} W_{k}}{\delta J_{c} (z) \,\delta J_{b} (y)} ~,}
can be expressed in terms of the $k$-dependent generating functional $W_{k}$. That way, we may rewrite \eqref{Rel1} as
\eqna{&& \hspace{-10pt} \delta^{(d+1)}(x-y) \delta_{a b} \\ &=& \int_{z} \left\{ \bigg( - \frac{\delta^{2} \Gamma_{k}}{\delta {\tilde \phi}_{a} (x) \,\delta \phi_{c} (z)}
- R^{\rmR}_{k, a c} (x,z) \bigg) \, \frac{\delta^{2} W_{k}}{\delta \tilde{J}_{c} (z) \,\delta J_{b} (y)} \right. \\
&& 
\left. -\: \frac{\delta^{2} \Gamma_{k}}{\delta {\tilde \phi}_{a} (x) \delta {\tilde \phi}_{c} (z)}
 \frac{\delta^{2} W_{k}}{\delta J_{c} (z) \, \delta J_{b} (y)} \right\}~.
\label{Rel2} \IEEEyesnumber}
With \eqref{eq:connectedtwopoint} this can be written in a compact matrix form,
\eqn{\left( \Gamma_{k}^{{\tilde \phi}\phi} + R^{\rmR}_{k} \right) G^{\rmR}_{k} + \Gamma^{{\tilde \phi}{\tilde \phi}}_{k}  i \tilde F_{k} = - \mathds{1} ~,}
where we have used the notation
\eqn{\Gamma^{{\tilde \phi} \phi}_{k, a b} (x,y) \equiv \frac{\delta^{2} \Gamma_{k} [\phi,{\tilde \phi}]}{\delta {\tilde \phi}_{a} (x) \, \delta \phi_{b} (y)} ~,}
for the functional derivatives of the effective average action. 

Starting from the remaining three identities for the functional derivatives of the sources, that is
\eqn{\frac{\delta \tilde{J}_{a} (x)}{\delta \tilde{J}_{b} (y)} = \delta^{(d+1)} (x-y) \delta_{a b} ~,}
and
\eqn{\frac{\delta J_{a} (x)}{\delta \tilde{J}_{b} (y)} = 0~, \quad \frac{\delta \tilde{J}_{a} (x)}{\delta J_{b} (y) } = 0~,}
we obtain the set of equations
\eqna{\left( \Gamma^{\phi {\tilde \phi}}_{k} + R^{\rmA}_{k} \right) G^{\rmA}_{k} + \Gamma_{k}^{\phi\phi} i F_{k} &=& - \mathds{1} ~,\\
\left( \Gamma_{k}^{{\tilde \phi}\phi} + R^{\rmR}_{k} \right) i F_{k} + \Gamma^{{\tilde \phi} {\tilde \phi}}_{k} G^{\rmA}_{k} &=& 0 ~,\\
\left( \Gamma^{\phi {\tilde \phi}}_{k} + R^{\rmA}_{k} \right) i \tilde F_{k} + \Gamma_{k}^{\phi\phi} G^{\rmR}_{k} &=& 0 ~.}
Together with our result from above, this linear system can be solved for the propagators $G^{\rmR,\rmA}_{k}, F_{k}$, and ${\tilde F}_{k}$:
\eqna{G_{k}^{\rmR} &=& - \left[ \left( \Gamma_{k}^{{\tilde \phi}\phi} + R_{k}^{\rmR} \right) - 
\Gamma_{k}^{{\tilde \phi}{\tilde \phi}} \left(\Gamma_{k}^{\phi{\tilde \phi}} + R_{k}^{\rmA}\right)^{-1}
\Gamma_{k}^{\phi\phi} \right]^{-1} , \\
G_{k}^{\rmA} &=& - \left[ \left( \Gamma_{k}^{\phi{\tilde \phi}} + R_{k}^{\rmA} \right) - 
\Gamma_{k}^{\phi \phi} \left(\Gamma_{k}^{{\tilde \phi}\phi} + R_{k}^{\rmR}\right)^{-1}
\Gamma_{k}^{{\tilde \phi}{\tilde \phi}} \right]^{-1} , \\
i F_{k} &=& - \left[ \Gamma_{k}^{\phi\phi} - \left( \Gamma_{k}^{\phi{\tilde \phi}} + R_{k}^{\rmA} \right) \Big( \Gamma_{k}^{{\tilde \phi}{\tilde \phi}} \Big)^{-1}
\left(\Gamma_{k}^{{\tilde \phi}\phi} + R_{k}^{\rmR}\right)\right]^{-1} , \\
i \tilde F_{k} &=& - \left[ \Gamma_{k}^{{\tilde \phi}{\tilde \phi}} - 
\left( \Gamma_{k}^{{\tilde \phi}\phi} + R_{k}^{\rmR} \right) \left( \Gamma_{k}^{\phi\phi} \right)^{-1}
\left(\Gamma_{k}^{\phi{\tilde \phi}} + R_{k}^{\rmA}\right)\right]^{-1} , 
\\ \IEEEyesnumber
\label{eq:relations}}
where the propagators depend on the field expectation values 
$\phi_{a}$ and ${\tilde \phi}_{a}$, i.e.\ $G_{k}^{\rmR,\rmA} = G_{k}^{\rmR,\rmA} [\phi,{\tilde \phi}]$, etc. We emphasize that the above is valid for the choice $R^{F,{\tilde F}}=0$, and
for nonvanishing $R^{F,{\tilde F}}$ the propagators take a different form \cite{Berges:2008sr}).

\subsection{Diagrammatics}
\vskip\pmsk

From the flow equation for the effective average action \eqref{FlowEquation}, we can construct the flow equations for arbitrary $n$-point functions by functional differentiation. As an example, we consider 
\eqna{{\dot \Gamma}^{{\tilde \phi}{\tilde \phi}}_{k,a b} (x,y) &=& - \frac{i}{2} 
\Tr \left\{ \frac{\delta^{2} G^{\rmR}_{k}[\phi,\tilde{\phi}]}{\delta {\tilde \phi}_{a} (x) \,\delta {\tilde \phi}_{b} (y)} {\dot R}^{\rmR}_{k} \right. \\
&& 
\left. +\: \frac{\delta^{2} G^{\rmA}_{k}[\phi,\tilde{\phi}]}{\delta {\tilde \phi}_{a} (x) \, \delta {\tilde \phi}_{b} (y)}{\dot R}^{\rmA}_{k} \right\}~,}
involving functional derivatives of the retarded and advanced propagators. For the retarded propagator we have, e.g.\
\eqna{\hspace{-2pt}\frac{\delta^{2} G^{\rmR}_{k,a b} [\phi,{\tilde \phi}]}{\delta {\tilde \phi}_{c}\, \delta {\tilde \phi}_{d}} &=& G^{\rmR}_{k,a e} \Gamma^{{\tilde \phi}\phi{\tilde \phi}{\tilde \phi}}_{k, e f c d} G^{\rmR}_{k,f b} \\
&& -\: G^{\rmR}_{k, a c} \Gamma^{{\tilde \phi}{\tilde \phi}{\tilde \phi}{\tilde \phi}}_{k, e f c d} \left(\Gamma^{\phi{\tilde{\phi}}}_{k} + R^{\rmA}_{k}\right)^{-1}_{f g} \Gamma^{\phi\phi}_{k, g h} G^{\rmR}_{k, h b} \\
&& +\: G^{\rmR}_{k, a e} \Gamma^{{\tilde \phi}{\tilde \phi}}_{k, e f} \left( \Gamma^{\phi{\tilde \phi}}_{k} + R^{\rmA}_{k} \right)^{-1}_{f g} \Gamma^{\phi{\tilde \phi}{\tilde \phi}{\tilde \phi}}_{k, g h c d} \\
&& \times \: \left( \Gamma^{\phi{\tilde \phi}}_{k} + R^{\rmA}_{k} \right)^{-1}_{k, h i} \Gamma^{\phi\phi}_{k, i j} G^{\rmR}_{k, j b} \\
&& -\: G^{\rmR}_{k, a e} \Gamma^{{\tilde \phi}{\tilde \phi}}_{k, e f} \left(\Gamma^{\phi{\tilde \phi}}_{k} + R^{\rmA}_{k} \right)^{-1}_{f g} \Gamma^{\phi\phi{\tilde \phi}{\tilde \phi}}_{k, g h c d} G^{\rmR}_{k, h b} \\
&& +\: \textrm{plus terms involving three-vertices} ~.}
This seems rather complicated at first sight. However, after taking derivatives, we can set the sources $\tilde{J} = J = 0$ to zero, which corresponds to evaluating all expressions at the field configuration $(\phi=\phi_0(k),\tilde{\phi}=0)$ that extremize the effective average action. Here $\phi_{0}(k=0)$ can be nonzero in the case of spontaneous symmetry breaking. According to the equation of motions, this configuration fulfills with \eqref{EoM1} for $\tilde{J} = J = 0$ (and $R_k^{F,{\tilde F}}=0$):
\eqn{\left. \Gamma_{k}^{\phi}\right|_{\phi=\phi_0(k),\tilde{\phi}=0} (x) = 0 ~.}
More generally, all the functional derivatives of the effective average action with respect to $\phi$ vanish at the extremum \cite{Chou:1984es}, i.e.
\eqn{\left. \frac{\delta^{n} \Gamma_{k} [\phi,{\tilde \phi}]}{\delta\phi(x_{1}) \, \delta\phi(x_{2}) \cdots \,\delta \phi(x_{n})} \right|_{\phi_0(k) ,{\tilde \phi} = 0} = 0 ~.}

In particular, we have $\Gamma^{\phi\phi}_{k} [\phi=\phi_0(k) ,{\tilde \phi} = 0] = 0$ so that the
\eqref{eq:relations} simplify considerably. The anomalous statistical propagator ${\tilde F}_{k}[\phi=\phi_0(k) ,{\tilde \phi} = 0]$ is zero and the nonvanishing propagators are given by
\eqna{G^{\rmR}_{k,a b} &=& - \left( \Gamma^{{\tilde \phi}\phi}_{k} + R^{\rmR}_{k} \right)^{-1}_{a b} =
\begin{picture}(40,21) (342,-198)
    \Text(344,-193)[lb]{$a$}
    \Text(371,-193)[lb]{$b$}
    \Text(383,-197)[lb]{,}
    \SetWidth{0.5}
    \SetColor{Black}
    \SetWidth{1.0}
    \Line(342,-195)(356,-195)
    \DashLine(356,-195)(376,-195){2}
\end{picture}\\
G^{\rmA}_{k, a b} &=& - \left( \Gamma^{\phi{\tilde \phi}}_{k} + R^{\rmA}_{k} \right)^{-1}_{a b} =
\begin{picture}(40,23) (342,-198)
    \SetWidth{0.5}
    \SetColor{Black}
    \Text(344,-193)[lb]{$a$}
    \Text(371,-193)[lb]{$b$}
    \Text(383,-197)[lb]{,}
    \SetWidth{1.0}
    \DashLine(342,-195)(356,-195){2}
    \Line(358,-195)(376,-195)
  \end{picture}\\
i F_{k, a b} &=& ~\left( G^{\rmR}_{k} \Gamma^{{\tilde \phi}{\tilde \phi}}_{k} G^{\rmA}_{k} \right)_{a b} \, =
\begin{picture}(40,23) (342,-198)
    \SetWidth{0.5}
    \SetColor{Black}
    \Text(344,-193)[lb]{$a$}
    \Text(371,-193)[lb]{$b$}
    \Text(383,-197)[lb]{,}
    \SetWidth{1.0}
    \Line(342,-195)(376,-195)
\end{picture}}
where we have also introduced their diagrammatic representation. 

In the following, for simplicity we will work in the symmetric phase where the macroscopic field expectation value vanishes, i.e.\ $\phi_{0}(k) = 0$. As a consequence, also all three-vertices vanish for the considered theory with interaction \eqref{InteractionPartAction} and using our above example we get the comparably compact expression:
\eqna{\left. \frac{\delta^{2} G^{\rmR}_{k,a b} [\phi,{\tilde \phi}]}{\delta {\tilde \phi}_{c} (x) \delta {\tilde \phi}_{d} (y)} \right|_{\phi,{\tilde \phi} = 0} &=& G^{\rmR}_{k, a e} \Gamma^{{\tilde \phi}\phi{\tilde \phi}{\tilde \phi}}_{k, e f c d} G^{\rmR}_{k, f b} \\
&& +\: i F_{k, a g} \Gamma^{\phi\phi{\tilde \phi}{\tilde \phi}}_{k, g h c d} G^{\rmR}_{k, h b}  ~.}

We will use a diagrammatic representations, where the retarded and advanced cutoff functions $R_{k}^{\rmR,\rmA}$ are denoted by the insertion of a cross, i.e.\
\eqna{
&&\begin{picture}(320,20) (28,-201)
    \SetWidth{0.5}
    \SetColor{Black}
    \Text(83,-198)[lb]{$\displaystyle {\dot R}_{k,ab}^{\rmR} \,=\, $}
\Text(125,-188)[lb]{$a$}
\Text(170,-188)[lb]{$b$}
\Text(179,-195)[lb]{~\,,}
    \SetWidth{1.0}
    \Line(145,-193)(151,-187)\Line(145,-187)(151,-193)
    \DashLine(151,-190)(124,-190){2}
    \Line(151,-190)(175,-190)
\end{picture} \\
&& \begin{picture}(320,20) (136,-194)
    \SetWidth{0.5}
    \SetColor{Black}
    \Text(191,-198)[lb]{$\displaystyle {\dot R}_{k,ab}^{\rmA} \,=\, $}
\Text(233,-188)[lb]{$a$}
\Text(278,-188)[lb]{$b$}
\Text(290,-195)[lb]{~\,,}
\SetWidth{1.0}
    \Line(256,-193)(262,-187)\Line(256,-187)(262,-193)
    \Line(259,-190)(232,-190)
    \DashLine(259,-190)(286,-190){2}
\end{picture}}
and the proper vertices, indicated by the full dot, are given by
\eqn{\begin{picture}(20,29) (280,-158)
    \SetWidth{0.5}
    \SetColor{Black}
\Text(196,-153)[lb]{$\displaystyle \Gamma_{k,abcd}^{{\tilde \phi}\phi\phi\phi} \,=\, $}
\Text(392,-150)[lb]{}
\Text(244,-135)[lb]{$c$}
\Text(273,-135)[lb]{$d$}
\Text(243,-161)[lb]{$a$}
\Text(273,-161)[lb]{$b$}
\Text(280,-152)[lb]{~,}
    \Text(300,-153)[lb]{$\displaystyle \Gamma_{k,abcd}^{{\tilde \phi}{\tilde \phi}\phi\phi} \,=\, $}
    \Text(350,-135)[lb]{$c$}
    \Text(378,-135)[lb]{$d$}
    \Text(349,-161)[lb]{$a$}
    \Text(378,-161)[lb]{$b$}
\SetWidth{1.0}
    \Vertex(260,-145){3.4}
    \Line(250,-135)(270,-155)
    \Line(260,-145)(270,-135)
    \DashLine(260,-145)(250,-155){2}
\SetWidth{1.0}
    \Vertex(365,-145){3.4}
    \Line(355,-135)(365,-145)
    \Line(365,-146)(375,-135)
    \DashLine(365,-145)(355,-155){2}
    \DashLine(375,-155)(365,-145){2}
\end{picture}}
and equivalently for the remaining four-vertices.

For the exact flow equation for the effective average action the one-loop form is written diagrammatically as
\eqn{
\begin{picture}(73,26) (24,-167)
    \SetWidth{0.5}
    \SetColor{Black}
    \Text(-6,-167)[lb]{$\displaystyle {\dot \Gamma}_k \,=\, -\frac{i}{2} \,\bigg\{ $}
    \Text(75,-160)[lb]{$\displaystyle +$}
    \Text(118,-167)[lb]{$\displaystyle \bigg\}~.$}
    \SetWidth{1.0}
    \Line(53,-148)(59,-142)\Line(53,-142)(59,-148)
    \CArc(56,-155)(10,90,270)
    \DashCArc(56,-155)(10,-90,90){2}
\SetWidth{1.0}
    \Line(98,-148)(104,-142)\Line(98,-142)(104,-148)
    \DashCArc(101,-155)(10,90,270){2}
    \CArc(101,-155)(10,-90,90)
\end{picture}}
Similarly, using a compact notation for the derivatives 
the flow equation for the two-point function takes the form
\eqna{
&&
\begin{picture}(73,34) (74,-54)
    \SetWidth{0.5}
    \SetColor{Black}
    \Text(127,-39)[lb]{$+$}
    \Text(26,-48)[lb]{$\displaystyle 
    {\dot \Gamma}_{k,ab}^{(2)} 
    \,\, =\,\, -\frac{i}{2} \, \bigg\{$}
  \Text(95,-55)[lb]{$a$}
  \Text(111,-55)[lb]{$b$}
  \Text(145,-55)[lb]{$a$}
  \Text(161,-55)[lb]{$b$}
    \SetWidth{1.0}
    \CArc(105,-34)(10,100,180)
    \DashCArc(105,-34)(10,180,-80){2}
    \DashCArc(105,-34)(10,0,100){2}
    \CArc(105,-34)(10,-80,0)
    \SetWidth{0.5}
    \Vertex(105,-44){3.4}
    \SetWidth{1.0}
    \Line(103,-21)(108,-27)\Line(103,-27)(108,-21)
    \Line(95,-45)(115,-45)
    \CArc(155,-34)(10,90,270)
    \DashCArc(155,-34)(10,0,90){2}
    \CArc(155,-34)(10,-90,0)
   \SetWidth{0.5}
    \Vertex(155,-44){3.4}
    \SetWidth{1.0}
    \Line(152,-21)(158,-27)\Line(152,-27)(158,-21)
    \Line(145,-45)(165,-45)
\end{picture} \\
&&\begin{picture}(73,34) (74,-54)
    \SetWidth{0.5}
    \SetColor{Black}
    \Text(79,-39)[lb]{$+$}
    \Text(127,-39)[lb]{$+$}
    \Text(178,-48)[lb]{$\displaystyle \bigg\} ~,$}
  \Text(95,-55)[lb]{$a$}
  \Text(111,-55)[lb]{$b$}
  \Text(145,-55)[lb]{$a$}
  \Text(161,-55)[lb]{$b$}
  \SetWidth{1.0}
    \DashCArc(105,-34)(10,90,180){2}
    \CArc(105,-34)(10,180,270)
    \CArc(105,-34)(10,0,90)
    \DashCArc(105,-34)(10,-90,0){2}
    \SetWidth{0.5}
    \Vertex(105,-44){3.4}
    \SetWidth{1.0}
    \Line(102,-21)(108,-27)\Line(102,-27)(108,-21)
    \Line(95,-45)(115,-45)
    \CArc(155,-34)(10,180,270)
    \DashCArc(155,-34)(10,90,180){2}
    \CArc(155,-34)(10,-90,90)
    \SetWidth{0.5}
    \Vertex(155,-44){3.4}
    \SetWidth{1.0}
    \Line(152,-21)(158,-27)\Line(152,-27)(158,-21)
    \Line(165,-45)(145,-45)
\end{picture}
}
and for the four-point function we have 
\eqna{
&&
\begin{picture}(349,18) (52,-52)
    \SetWidth{0.5}
    \SetColor{Black}
    \Text(50,-50)[lb]{$\displaystyle 
    {\dot \Gamma}_{k,abcd}^{(4)} \, = $}
\end{picture} \\
&&
\begin{picture}(349,53) (60,-75)
    \SetWidth{0.5}
    \SetColor{Black}
    \Text(124,-42)[lb]{$\displaystyle +$}
    \Text(174,-42)[lb]{$\displaystyle +$}
    \Text(226,-42)[lb]{$\displaystyle +$}
    \Text(60,-50)[lb]{$ -\frac{i}{8} \, \bigg\{$}
    \Text(83,-60)[lb]{$a$}
    \Text(96,-60)[lb]{$b$}
    \Text(105,-60)[lb]{$c$}
    \Text(115,-60)[lb]{$d$}
    \Text(133,-60)[lb]{$a$}
    \Text(145,-60)[lb]{$b$}
    \Text(155,-60)[lb]{$c$}
    \Text(165,-60)[lb]{$d$}
    \Text(183,-60)[lb]{$a$}
    \Text(195,-60)[lb]{$b$}
    \Text(205,-60)[lb]{$c$}
    \Text(215,-60)[lb]{$d$}
    \Text(233,-60)[lb]{$a$}
    \Text(245,-60)[lb]{$b$}
    \Text(255,-60)[lb]{$c$}
    \Text(265,-60)[lb]{$d$}
    \SetWidth{1.0}
    \CArc(102,-34)(10,90,140)
    \DashCArc(102,-34)(10,140,-80){2}
    \DashCArc(102,-34)(10,20,90){2}
    \CArc(102,-34)(10,-80,20)
    \SetWidth{0.5}
    \Vertex(94,-39){3.4}
    \Vertex(110,-39){3.4}
    \SetWidth{1.0}
    \Line(99,-21)(105,-27)\Line(99,-27)(105,-21)
    \Line(94,-39)(87,-51)
    \Line(94,-39)(97,-51)
    \Line(110,-39)(107,-51)
    \Line(110,-39)(117,-51)
    \CArc(152,-34)(10,90,140)
    \DashCArc(152,-34)(10,20,90){2}
    \DashCArc(152,-34)(10,140,210){2}
    \CArc(152,-34)(10,-30,20)
    \CArc(152,-34)(10,210,260)
    \DashCArc(152,-34)(10,260,-30){2}
    \SetWidth{0.5}
    \Vertex(144,-39){3.4}
    \Vertex(160,-39){3.4}
    \SetWidth{1.0}
    \Line(149,-21)(155,-27)\Line(149,-27)(155,-21)
    \Line(144,-39)(137,-51)
    \Line(144,-39)(147,-51)
    \Line(160,-39)(157,-51)
    \Line(160,-39)(167,-51)
    \CArc(202,-34)(10,90,140)
    \DashCArc(202,-34)(10,20,90){2}
    \DashCArc(202,-34)(10,140,210){2}
    \CArc(202,-34)(10,210,20)
    \SetWidth{0.5}
    \Vertex(194,-39){3.4}
    \Vertex(210,-39){3.4}
    \SetWidth{1.0}
    \Line(199,-21)(205,-27)\Line(199,-27)(205,-21)
    \Line(194,-39)(187,-51)
    \Line(194,-39)(197,-51)
    \Line(210,-39)(207,-51)
    \Line(210,-39)(217,-51)
\SetWidth{1.0}
    \CArc(252,-34)(10,90,210)
    \DashCArc(252,-34)(10,210,280){2}
    \DashCArc(252,-34)(10,20,90){2}
    \CArc(252,-34)(10,-80,20)
    \SetWidth{0.5}
   \Vertex(244,-39){3.4}
    \Vertex(260,-39){3.4}
    \SetWidth{1.0}
    \Line(249,-21)(255,-27)\Line(249,-27)(255,-21)
    \Line(244,-39)(237,-51)
    \Line(244,-39)(247,-51)
    \Line(260,-39)(257,-51)
    \Line(260,-39)(267,-51)
\end{picture}\\
&&  
\begin{picture}(349,48) (68,-75)
    \SetWidth{0.5}
    \SetColor{Black}
    \Text(83,-42)[lb]{$\displaystyle +$}
    \Text(133,-42)[lb]{$\displaystyle +$}
    \Text(183,-42)[lb]{$\displaystyle +$}
    \Text(234,-42)[lb]{$\displaystyle +$}
    \Text(91,-60)[lb]{$a$}
    \Text(103,-60)[lb]{$b$}
    \Text(113,-60)[lb]{$c$}
    \Text(123,-60)[lb]{$d$}
    \Text(141,-60)[lb]{$a$}
    \Text(153,-60)[lb]{$b$}
    \Text(163,-60)[lb]{$c$}
    \Text(173,-60)[lb]{$d$}
    \Text(191,-60)[lb]{$a$}
    \Text(203,-60)[lb]{$b$}
    \Text(213,-60)[lb]{$c$}
    \Text(223,-60)[lb]{$d$}
    \Text(241,-60)[lb]{$a$}
    \Text(253,-60)[lb]{$b$}
    \Text(263,-60)[lb]{$c$}
    \Text(273,-60)[lb]{$d$}
    \SetWidth{1.0}
    \CArc(110,-34)(10,90,260)
    \DashCArc(110,-34)(10,260,330){2}
    \DashCArc(110,-34)(10,20,90){2}
    \CArc(110,-34)(10,-30,20)
    \SetWidth{0.5}
    \Vertex(102,-39){3.4}
    \Vertex(118,-39){3.4}
    \SetWidth{1.0}
    \Line(107,-21)(113,-27)\Line(107,-27)(113,-21)
    \Line(102,-39)(95,-51)
    \Line(102,-39)(105,-51)
    \Line(118,-39)(115,-51)
    \Line(118,-39)(125,-51)
    \CArc(160,-34)(10,90,140)
    \DashCArc(160,-34)(10,20,90){2}
    \CArc(160,-34)(10,90,20)
    \SetWidth{0.5}
    \Vertex(152,-39){3.4}
    \Vertex(168,-39){3.4}
    \SetWidth{1.0}
    \Line(157,-21)(163,-27)\Line(157,-27)(163,-21)
    \Line(152,-39)(145,-51)
    \Line(152,-39)(155,-51)
    \Line(168,-39)(165,-51)
    \Line(168,-39)(175,-51)
    \CArc(210,-34)(10,40,90)
    \DashCArc(210,-34)(10,-30,40){2}
    \DashCArc(210,-34)(10,90,160){2}
    \CArc(210,-34)(10,160,210)
    \DashCArc(210,-34)(10,210,280){2}
    \CArc(210,-34)(10,-80,-30)
    \SetWidth{0.5}
    \Vertex(202,-39){3.4}
    \Vertex(218,-39){3.4}
    \SetWidth{1.0}
    \Line(207,-21)(213,-27)\Line(207,-27)(213,-21)
    \Line(202,-39)(195,-51)
    \Line(202,-39)(205,-51)
    \Line(218,-39)(215,-51)
    \Line(218,-39)(225,-51)
    \SetWidth{1.0}
    \CArc(260,-34)(10,40,90)
    \DashCArc(260,-34)(10,260,40){2}
    \CArc(260,-34)(10,160,260)
    \DashCArc(260,-34)(10,90,160){2}
    \SetWidth{0.5}
    \Vertex(252,-39){3.4}
    \Vertex(268,-39){3.4}
    \SetWidth{1.0}
    \Line(257,-21)(263,-27)\Line(257,-27)(263,-21)
    \Line(252,-39)(245,-51)
    \Line(252,-39)(255,-51)
    \Line(268,-39)(265,-51)
    \Line(268,-39)(275,-51)
\end{picture}\\
&&
\begin{picture}(349,48) (68,-75)
    \SetWidth{0.5}
    \SetColor{Black}
    \Text(133,-42)[lb]{$\displaystyle +$}
    \Text(183,-42)[lb]{$\displaystyle +$}
    \Text(234,-42)[lb]{$\displaystyle +$}
    \Text(83,-42)[lb]{$\displaystyle +$}
    \Text(91,-60)[lb]{$a$}
    \Text(103,-60)[lb]{$b$}
    \Text(113,-60)[lb]{$c$}
    \Text(123,-60)[lb]{$d$}
    \Text(141,-60)[lb]{$a$}
    \Text(153,-60)[lb]{$b$}
    \Text(163,-60)[lb]{$c$}
    \Text(173,-60)[lb]{$d$}
    \Text(191,-60)[lb]{$a$}
    \Text(203,-60)[lb]{$b$}
    \Text(213,-60)[lb]{$c$}
    \Text(223,-60)[lb]{$d$}
    \Text(241,-60)[lb]{$a$}
    \Text(253,-60)[lb]{$b$}
    \Text(263,-60)[lb]{$c$}
    \Text(273,-60)[lb]{$d$}
    \SetWidth{1.0}
    \CArc(110,-34)(10,40,90)
    \DashCArc(110,-34)(10,-30,40){2}
    \CArc(110,-34)(10,160,-30)
    \DashCArc(110,-34)(10,90,160){2}
    \SetWidth{0.5}
    \Vertex(102,-39){3.4}
    \Vertex(118,-39){3.4}
    \SetWidth{1.0}
    \Line(107,-21)(113,-27)\Line(107,-27)(113,-21)
    \Line(102,-39)(95,-51)
    \Line(102,-39)(105,-51)
    \Line(118,-39)(115,-51)
    \Line(118,-39)(125,-51)
\SetWidth{1.0}
    \CArc(160,-34)(10,-80,90)
    \DashCArc(160,-34)(10,90,160){2}
    \CArc(160,-34)(10,160,210)
    \DashCArc(160,-34)(10,210,280){2}
    \SetWidth{0.5}
   \Vertex(152,-39){3.4}
    \Vertex(168,-39){3.4}
    \SetWidth{1.0}
    \Line(157,-21)(163,-27)\Line(157,-27)(163,-21)
    \Line(152,-39)(145,-51)
    \Line(152,-39)(155,-51)
    \Line(168,-39)(165,-51)
    \Line(168,-39)(175,-51)
    \CArc(210,-34)(10,-30,90)
    \DashCArc(210,-34)(10,260,-30){2}
    \CArc(210,-34)(10,160,260)
    \DashCArc(210,-34)(10,90,160){2}
    \SetWidth{0.5}
    \Vertex(202,-39){3.4}
    \Vertex(218,-39){3.4}
    \SetWidth{1.0}
    \Line(207,-21)(213,-27)\Line(207,-27)(213,-21)
    \Line(202,-39)(195,-51)
    \Line(202,-39)(205,-51)
    \Line(218,-39)(215,-51)
    \Line(218,-39)(225,-51)
    \CArc(260,-34)(10,160,90)
    \DashCArc(260,-34)(10,90,160){2}
    \Vertex(252,-39){3.4}
    \Vertex(268,-39){3.4}
    \SetWidth{1.0}
    \Line(257,-21)(263,-27)\Line(257,-27)(263,-21)
    \Line(252,-39)(245,-51)
    \Line(252,-39)(255,-51)
    \Line(268,-39)(265,-51)
    \Line(268,-39)(275,-51)
\end{picture}\\
&&
\begin{picture}(349,28) (65,-56)
    \SetWidth{0.5}
    \SetColor{Black}
 \Text(80,-42)[lb]{$\displaystyle + \quad P(a,b,c,d) \,\, \bigg\}$}
\end{picture}\\
&&
\begin{picture}(349,30) (72,-61)
    \SetWidth{0.5}
    \SetColor{Black}
    \Text(130,-44)[lb]{$\displaystyle +$}
    \Text(180,-44)[lb]{$\displaystyle +$}
    \Text(230,-45)[lb]{$\displaystyle +$}
    \Text(280,-50)[lb]{$\bigg\}~.$}
    \Text(73,-50)[lb]{$\displaystyle - \frac{i}{2}\, \bigg\{$}
\Text(93,-63)[lb]{$a$}
\Text(102,-63)[lb]{$b$}
\Text(113,-63)[lb]{$c$}
\Text(122,-63)[lb]{$d$}
\Text(143,-63)[lb]{$a$}
\Text(152,-63)[lb]{$b$}
\Text(163,-63)[lb]{$c$}
\Text(172,-63)[lb]{$d$}
\Text(193,-63)[lb]{$a$}
\Text(202,-63)[lb]{$b$}
\Text(213,-63)[lb]{$c$}
\Text(222,-63)[lb]{$d$}
\Text(243,-63)[lb]{$a$}
\Text(252,-63)[lb]{$b$}
\Text(263,-63)[lb]{$c$}
\Text(272,-63)[lb]{$d$}
\SetWidth{1.0}
    \CArc(110,-34)(10,90,180)
    \DashCArc(110,-34)(10,180,270){2}
    \DashCArc(110,-34)(10,0,90){2}
    \CArc(110,-34)(10,-90,0)
    \SetWidth{0.5}
    \Vertex(110,-44){3.4}
    \SetWidth{1.0}
    \Line(107,-21)(113,-27)\Line(107,-27)(113,-21)
    \Line(97,-55)(110,-44)
    \Line(105,-55)(110,-44)
    \Line(115,-55)(110,-44)
    \Line(123,-55)(110,-44)
    \CArc(160,-34)(10,90,270)
    \DashCArc(160,-34)(10,0,90){2}
    \CArc(160,-34)(10,-90,0)
   \SetWidth{0.5}
    \Vertex(160,-44){3.4}
    \SetWidth{1.0}
    \Line(157,-21)(163,-27)\Line(157,-27)(163,-21)
    \Line(147,-55)(160,-44)
    \Line(155,-55)(160,-44)
    \Line(165,-55)(160,-44)
    \Line(173,-55)(160,-44)
    \DashCArc(210,-34)(10,90,180){2}
    \CArc(210,-34)(10,180,270)
    \CArc(210,-34)(10,0,90)
    \DashCArc(210,-34)(10,-90,0){2}
    \SetWidth{0.5}
    \Vertex(210,-44){3.4}
    \SetWidth{1.0}
    \Line(207,-21)(213,-27)\Line(207,-27)(213,-21)
    \Line(197,-55)(210,-44)
    \Line(205,-55)(210,-44)
    \Line(215,-55)(210,-44)
    \Line(223,-55)(210,-44)
    \CArc(260,-34)(10,180,90)
    \DashCArc(260,-34)(10,90,180){2}
   \SetWidth{0.5}
    \Vertex(260,-44){3.4}
    \SetWidth{1.0}
    \Line(257,-21)(263,-27)\Line(257,-27)(263,-21)
    \Line(247,-55)(260,-44)
    \Line(255,-55)(260,-44)
    \Line(265,-55)(260,-44)
    \Line(273,-55)(260,-44)
\end{picture}}
Here, $P(a,b,c,d)$ denotes all possible permutations of the indices on legs of the respective diagrams.
The diagrammatic representation of the flow equations for $n$-point functions follows the standard construction rules: draw all combinations of propagators $G^{\rmR,\rmA}_{k}$ and $F_{k}$, and vertices $\Gamma^{(n)}_{k}$ and attach the appropriate symmetry factors, taking into account that certain diagrams may either vanish or are identical. We emphasize, however, again that in the diagrammatic representation the propagators are evaluated at an extremum of the effective average action. 

We may simplify the construction rules of the flow equations even further by introducing the derivative operator $\tilde{\partial}_{k}$,
\eqnlabel{
\tilde{\partial}_{k} \Gamma_{k} [\phi,{\tilde \phi}] \equiv \Tr \left[ {\dot R}^{\rmR}_{k} \frac{\delta}{\delta R^{\rmR}_{k}} + {\dot R}^{\rmA}_{k} \frac{\delta}{\delta R^{\rmA}_{k}} \right] \Gamma_{k}~.
\label{eq:tildedel}}
With this we may reduce expressions like
\eqn{\left( \Gamma^{{\tilde \phi}\phi}_{k} + R^{\rmR}_{k} \right)^{-1} {\dot R}^{\rmR}_{k} \left( \Gamma^{{\tilde \phi}\phi}_{k} + R^{\rmR}_{k} \right)^{-1} 
= - \tilde{\partial}_{k} \left( \Gamma^{{\tilde \phi}\phi} + R^{\rmR}_{k} \right)^{-1} ~,}
to a simple form. For instance, the flow equation for the two-point function can be written as
\eqn{
\begin{picture}(249,28) (15,-54)
    \SetWidth{0.5}
    \SetColor{Black}
    \Text(120,-42)[lb]{$\displaystyle +$}
    \Text(170,-42)[lb]{$\displaystyle +$}
    \Text(221,-47)[lb]{$\bigg\}~.$}
    \Text(16,-47)[lb]{$\displaystyle {\dot \Gamma}_{k,ab}^{(2)} 
    \, =\, -\, \frac{i}{2} \, {\tilde \partial}_k \,\, \bigg\{$}
  \Text(90,-56)[lb]{$a$}
  \Text(107,-56)[lb]{$b$}
  \Text(140,-56)[lb]{$a$}
  \Text(157,-56)[lb]{$b$}
  \Text(190,-56)[lb]{$a$}
  \Text(207,-56)[lb]{$b$}
    \SetWidth{1.0}
    \DashCArc(100,-34)(10,90,-90){2}
    \CArc(100,-34)(10,-90,90)
    \SetWidth{0.5}
    \Vertex(100,-44){3.4}
    \SetWidth{1.0}
    \Line(90,-45)(110,-45)
    \CArc(150,-34)(10,90,270)
    \DashCArc(150,-34)(10,-90,90){2}
   \SetWidth{0.5}
    \Vertex(150,-44){3.4}
    \SetWidth{1.0}
    \Line(140,-45)(160,-45)
   \SetWidth{1.0}
    \CArc(200,-34)(10,90,-90)
    \CArc(200,-34)(10,-90,90)
    \SetWidth{0.5}
    \Vertex(200,-44){3.4}
    \SetWidth{1.0}
    \Line(190,-45)(210,-45)
\end{picture}}
Equivalently, for the four-point function $\Gamma_{k}^{(4)}$ we have 
\eqna{
&&
\begin{picture}(249,26) (26,-50)
    \SetWidth{0.5}
    \SetColor{Black}
    \Text(27,-45)[lb]{$\displaystyle {\dot \Gamma}_{k,abcd}^{(4)} \, = \, $}
\end{picture}\\
&&
\begin{picture}(249,26) (30,-50)
    \SetWidth{0.5}
    \SetColor{Black}
    \Text(125,-37)[lb]{$\displaystyle +$}
    \Text(187,-37)[lb]{$\displaystyle +$}
    \Text(30,-45)[lb]{$  -\frac{i}{16} \, \tilde{\partial}_k\, \bigg\{$}
    \Text(72,-28)[lb]{$a$}
    \Text(72,-47)[lb]{$b$}
    \Text(117,-28)[lb]{$c$}
    \Text(117,-47)[lb]{$d$}
    \Text(135,-28)[lb]{$a$}
    \Text(135,-47)[lb]{$b$}
    \Text(180,-28)[lb]{$c$}
    \Text(180,-47)[lb]{$d$}
    \Text(196,-28)[lb]{$a$}
    \Text(196,-47)[lb]{$b$}
    \Text(241,-28)[lb]{$c$}
    \Text(241,-47)[lb]{$d$}
    \SetWidth{1.0}
    \CArc(97,-34)(10,-90,90)
    \DashCArc(97,-34)(10,90,270){2}
    \SetWidth{0.5}
    \Vertex(87,-34){3.4}
    \Vertex(107,-34){3.4}
    \SetWidth{1.0}
    \Line(87,-34)(79,-25)
    \Line(87,-34)(79,-43)
    \Line(107,-34)(115,-25)
    \Line(107,-34)(115,-43)
    \DashCArc(159,-34)(10,-90,0){2}
    \CArc(159,-34)(10,180,-90)
    \DashCArc(159,-34)(10,90,180){2}
    \CArc(159,-34)(10,0,90)
    \SetWidth{0.5}
    \Vertex(149,-34){3.4}
    \Vertex(169,-34){3.4}
    \SetWidth{1.0}
    \Line(149,-34)(141,-25)
    \Line(149,-34)(141,-43)
    \Line(169,-34)(177,-25)
    \Line(169,-34)(177,-43)
    \CArc(220,-34)(10,180,90)
    \DashCArc(220,-34)(10,90,180){2}
    \SetWidth{0.5}
    \Vertex(210,-34){3.4}
    \Vertex(230,-34){3.4}
    \SetWidth{1.0}
    \Line(210,-34)(202,-25)
    \Line(210,-34)(202,-43)
    \Line(230,-34)(238,-25)
    \Line(230,-34)(238,-43)
\end{picture}\\
&&  
\begin{picture}(249,34) (30,-51)
    \SetWidth{0.5}
    \SetColor{Black}
    \Text(125,-37)[lb]{$\displaystyle +$}
    \Text(187,-37)[lb]{$\displaystyle +$}
    \Text(58,-37)[lb]{$\displaystyle +$}
    \Text(72,-28)[lb]{$a$}
    \Text(72,-47)[lb]{$b$}
    \Text(117,-28)[lb]{$c$}
    \Text(117,-47)[lb]{$d$}
    \Text(135,-28)[lb]{$a$}
    \Text(135,-47)[lb]{$b$}
    \Text(180,-28)[lb]{$c$}
    \Text(180,-47)[lb]{$d$}
    \Text(196,-28)[lb]{$a$}
    \Text(196,-47)[lb]{$b$}
    \Text(241,-28)[lb]{$c$}
    \Text(241,-47)[lb]{$d$}
    \SetWidth{1.0}
    \DashCArc(97,-34)(10,0,90){2}
    \CArc(97,-34)(10,90,180)
    \DashCArc(97,-34)(10,180,270){2}
    \CArc(97,-34)(10,270,360)
    \SetWidth{0.5}
    \Vertex(87,-34){3.4}
    \Vertex(107,-34){3.4}
    \SetWidth{1.0}
    \Line(87,-34)(79,-25)
    \Line(87,-34)(79,-43)
    \Line(107,-34)(115,-25)
    \Line(107,-34)(115,-43)
    \DashCArc(159,-34)(10,-90,90){2}
    \CArc(159,-34)(10,90,270)
    \SetWidth{0.5}
    \Vertex(149,-34){3.4}
    \Vertex(169,-34){3.4}
    \SetWidth{1.0}
    \Line(149,-34)(141,-25)
    \Line(149,-34)(141,-43)
    \Line(169,-34)(177,-25)
    \Line(169,-34)(177,-43)
    \CArc(220,-34)(10,90,360)
    \DashCArc(220,-34)(10,0,90){2}
    \SetWidth{0.5}
    \Vertex(210,-34){3.4}
    \Vertex(230,-34){3.4}
    \SetWidth{1.0}
    \Line(210,-34)(202,-25)
    \Line(210,-34)(202,-43)
    \Line(230,-34)(238,-25)
    \Line(230,-34)(238,-43)
\end{picture}\\
&&  
\begin{picture}(249,34) (30,-50)
    \SetWidth{0.5}
    \SetColor{Black}
    \Text(125,-37)[lb]{$\displaystyle +$}
    \Text(187,-37)[lb]{$\displaystyle +$}
    \Text(58,-37)[lb]{$\displaystyle +$}
    \Text(72,-28)[lb]{$a$}
    \Text(72,-47)[lb]{$b$}
    \Text(117,-28)[lb]{$c$}
    \Text(117,-47)[lb]{$d$}
    \Text(135,-28)[lb]{$a$}
    \Text(135,-47)[lb]{$b$}
    \Text(180,-28)[lb]{$c$}
    \Text(180,-47)[lb]{$d$}
    \Text(196,-28)[lb]{$a$}
    \Text(196,-47)[lb]{$b$}
    \Text(241,-28)[lb]{$c$}
    \Text(241,-47)[lb]{$d$}
    \SetWidth{1.0}
    \CArc(97,-34)(10,-90,180)
    \DashCArc(97,-34)(10,180,270){2}
    \SetWidth{0.5}
    \Vertex(87,-34){3.4}
    \Vertex(107,-34){3.4}
    \SetWidth{1.0}
    \Line(87,-34)(79,-25)
    \Line(87,-34)(79,-43)
    \Line(107,-34)(115,-25)
    \Line(107,-34)(115,-43)
    \CArc(159,-34)(10,0,270)
    \DashCArc(159,-34)(10,270,360){2}
    \SetWidth{0.5}
    \Vertex(149,-34){3.4}
    \Vertex(169,-34){3.4}
    \SetWidth{1.0}
    \Line(149,-34)(141,-25)
    \Line(149,-34)(141,-43)
    \Line(169,-34)(177,-25)
    \Line(169,-34)(177,-43)
    \CArc(220,-34)(10,0,360)
    \SetWidth{0.5}
    \Vertex(210,-34){3.4}
    \Vertex(230,-34){3.4}
    \SetWidth{1.0}
    \Line(210,-34)(202,-25)
    \Line(210,-34)(202,-43)
    \Line(230,-34)(238,-25)
    \Line(230,-34)(238,-43)
\end{picture}\\
&&
\begin{picture}(249,24) (30,-48)
  \SetWidth{0.5}
  \SetColor{Black}
  \Text(58,-44)[lb]{$\displaystyle + \quad P(a,b,c,d) \, \bigg\}$}
\end{picture}
\nonumber\\
&&
\begin{picture}(249,32) (33,-58)
    \SetWidth{0.5}
    \SetColor{Black}
    \Text(115,-44)[lb]{$\displaystyle +$}
    \Text(175,-44)[lb]{$\displaystyle +$}
    \Text(232,-50)[lb]{$\bigg\}~.$}
    \Text(36,-50)[lb]{$\displaystyle - \frac{i}{2}\,\, \tilde{\partial}_k\,\, \bigg\{$}
\Text(73,-63)[lb]{$a$}
\Text(83,-63)[lb]{$b$}
\Text(93,-63)[lb]{$c$}
\Text(102,-63)[lb]{$d$}
\Text(133,-63)[lb]{$a$}
\Text(143,-63)[lb]{$b$}
\Text(153,-63)[lb]{$c$}
\Text(162,-63)[lb]{$d$}
\Text(193,-63)[lb]{$a$}
\Text(203,-63)[lb]{$b$}
\Text(213,-63)[lb]{$c$}
\Text(222,-63)[lb]{$d$}
\SetWidth{1.0}
    \CArc(90,-34)(10,-90,90)
    \DashCArc(90,-34)(10,90,270){2}
    \SetWidth{0.5}
    \Vertex(90,-44){3.4}
    \SetWidth{1.0}
    \Line(77,-55)(90,-44)
    \Line(85,-55)(90,-44)
    \Line(95,-55)(90,-44)
    \Line(103,-55)(90,-44)
    \CArc(150,-34)(10,90,270)
    \DashCArc(150,-34)(10,-90,90){2}
   \SetWidth{0.5}
    \Vertex(150,-44){3.4}
    \SetWidth{1.0}
    \Line(137,-55)(150,-44)
    \Line(145,-55)(150,-44)
    \Line(155,-55)(150,-44)
    \Line(163,-55)(150,-44)
    \CArc(210,-34)(10,90,270)
    \CArc(210,-34)(10,-90,90)
    \SetWidth{0.5}
    \Vertex(210,-44){3.4}
    \SetWidth{1.0}
    \Line(197,-55)(210,-44)
    \Line(205,-55)(210,-44)
    \Line(215,-55)(210,-44)
    \Line(223,-55)(210,-44)
\end{picture}}

\section{Solving truncated flow equations}
\label{sec:truncation}

By calculating the functional derivatives of the effective average action, we extracted above the flow equations for $n$-point functions. We will use these flow equations in the following to study the behavior in the vicinity of possible nonthermal fixed points.
Fixed points correspond to translationally invariant scaling solutions for $n$-point functions both in space and time. Therefore, we are interested in the limit $t_{0} \rightarrow - \infty$ for which the dependence on the details about the initial conditions encoded in $\varrho(t_0)$ are lost. As discussed in section \ref{sec:weakturbulence}, conserved quantities will play an important role and we will not impose a fluctuation-dissipation relation to be able to describe nonthermal fixed points.

\subsection{Stationarity condition}
\vskip\pmsk

The presence of nonthermal scaling solutions was discussed in section \ref{sec:weakturbulence} based on a stationarity condition for the nonequilibrium time evolution equations.
An equivalent (scale-dependent) stationarity condition can be obtained from the functional renormalization group, where it appears as a nontrivial identity relating the various second functional derivatives of $\Gamma_k[\phi,{\tilde \phi}]$. For spacetime translation invariant systems, it is very convenient to consider the correlation functions in Fourier space.
We start by writing down the following identity in momentum space,
\eqna{&& \hspace{-25pt} i \Gamma^{{\tilde \phi}{\tilde \phi}}_{k, a c} (p) \left\{ G^{\rmR}_{k, c a} (p) - G^{\rmA}_{k, c a} (p) \right\} \\
&=& -\: i G^{\rmR}_{k, a c} (p) \Gamma^{{\tilde \phi}{\tilde \phi}}_{k, c d} G^{\rmA}_{k, d e} (p) \\
&& \times\:\left\{ \left( G^{\rmR}_{k} \right)^{-1}_{e a} (p) - \left( G^{\rmA}_{k} \right)^{-1}_{e a} (p) \right\} ~.\IEEEyesnumber
\label{eq:identity1}} 
In order to interpret further the different combinations of terms appearing in this equation,
we write the two-point functions $\Gamma^{(2)}_{k}$ in the form
\eqna{\begin{pmatrix} \Gamma^{\phi \phi}_{k} & \Gamma^{\phi{\tilde \phi}}_{k} \\ \Gamma^{{\tilde \phi}\phi}_{k} & \Gamma^{{\tilde \phi}{\tilde \phi}}_{k} \end{pmatrix}
&=& \begin{pmatrix} 0 & - \left( G^{\rmA}_{k} \right)^{-1} - R^{\rmA}_{k} \\ - \left( G^{\rmR}_{k} \right)^{-1} - R^{\rmR}_{k} & \left( G^{\rmR}_{k} \right)^{-1} i F_{k}
\left( G^{\rmA}_{k} \right)^{-1} \end{pmatrix} \\
&\equiv& \begin{pmatrix} 0 & i D^{-1} \\ i D^{-1} & 0 \end{pmatrix} - \begin{pmatrix} 0 & \Sigma^{\rmA}_{k} \\ \Sigma^{\rmR}_{k} & i \Sigma^F_{k} \end{pmatrix} ~,}
where $i D^{-1}$ is the free inverse propagator. The second line defines the self-energies $\Sigma^{\rmR,\rmA,F}_{k}$, which in turn are given by
\eqna{\Sigma^F_{k, a b} &\equiv& i \Gamma^{{\tilde \phi}{\tilde \phi}}_{k, a b} ~, \\
\Sigma^{\rho}_{k, a b} &\equiv& \Sigma^{\rmR}_{k, a b} - \Sigma^{\rmA}_{k, a b} 
= \Gamma^{\phi {\tilde \phi}}_{k, a b} - \Gamma^{{\tilde \phi} \phi}_{k, ab} ~,}
where we have used that $R^{\rmR}_{k} = R^{\rmA}_{k}$ for the considered class of cutoff 
functions \eqref{eq:cutoffclass}. With these identifications, the identity \eqref{eq:identity1} can be written in terms of the self-energies as
\eqnlabel{\Sigma^F_{k, a b} (p)\, \rho_{k, b a } (p) - F_{k, a b} (p)\, \Sigma^{\rho}_{k, b a} (p) = 0~.
\label{Stationarity}}
This equation is well-known in nonequilibrium physics. In the language of Boltzmann dynamics employed in section \ref{sec:weakturbulence}, it essentially\footnote{Using the Wigner coordinates employed in section \ref{sec:weakturbulence}, the nonequilibrium time evolution of the statistical function is given by \cite{Berges:2005md}
\eqn{2p^\mu\partial_{X^\mu} F_{ab}(X,p) = i \left( \Sigma^F_{ac} \rho_{cb} - F_{ac} \Sigma^\rho_{cb}  \right)(X,p).}
The condition \eqref{Stationarity} is related to this by taking the trace at a stationary point, where the correlation functions become independent of $X$.
} states that `gain terms' equal `loss terms' for which stationarity is achieved~\cite{Berges:2004yj}. This aspect will be analyzed in detail in section \ref{sec:nonthermalfp}.

Of course, the condition \eqref{Stationarity} is trivially fulfilled if the fluc\-tu\-at\-ion-dissipation relation holds, i.e.\ in thermal equilibrium where
\eqna{F_{k}^{\rm (eq)} (p) &=& -i \left( n_{BE}(p^{0}) + \frac{1}{2} \right) \rho_{k}^{\rm (eq)} (p) ~, \\
\Sigma^{F \,{\rm (eq)}}_{k} (p) &=& -i \left( n_{BE}(p^{0}) + \frac{1}{2} \right) \Sigma^{\rho \,{\rm (eq)}}_{k} (p) ~,}
for the propagators and self-energies \cite{Berges:2004yj}. These relations will \emph{not} be assumed in the following. 

Using the representation of the self-energies in terms of two-point functions we may immediately write down the flow equations for the statistical component $\Sigma^{F}_{k}$,
which is given by
\eqna{
\begin{picture}(249,32) (40,-54)
    \SetWidth{0.5}
    \SetColor{Black}
    \Text(142,-42)[lb]{$\displaystyle +$}
    \Text(192,-42)[lb]{$\displaystyle +$}
    \Text(238,-47)[lb]{$\, \bigg\}~,$}
    \Text(40,-47)[lb]{$\displaystyle {\dot \Sigma}^{F}_{k, a b} \,\,=\,\,  \frac{1}{2} \, {\tilde \partial}_k \,\, \bigg\{ \,$}
  \Text(110,-56)[lb]{$a$}
  \Text(127,-56)[lb]{$b$}
  \Text(160,-56)[lb]{$a$}
  \Text(177,-56)[lb]{$b$}
  \Text(210,-56)[lb]{$a$}
  \Text(227,-56)[lb]{$b$}
    \SetWidth{1.0}
    \DashCArc(120,-34)(10,90,-90){2}
    \CArc(120,-34)(10,-90,90)
    \SetWidth{0.5}
    \Vertex(120,-44){3.4}
    \SetWidth{1.0}
    \DashLine(110,-45)(130,-45){2}
    \CArc(170,-34)(10,90,270)
    \DashCArc(170,-34)(10,-90,90){2}
   \SetWidth{0.5}
    \Vertex(170,-44){3.4}
    \SetWidth{1.0}
    \DashLine(160,-45)(180,-45){2}
   \SetWidth{1.0}
    \CArc(220,-34)(10,90,-90)
    \CArc(220,-34)(10,-90,90)
    \SetWidth{0.5}
    \Vertex(220,-44){3.4}
    \SetWidth{1.0}
    \DashLine(210,-45)(230,-45){2}
\end{picture}}
reading in momentum space
\eqna{{\dot \Sigma}^{F}_{k, a b} (p) &=& \frac{1}{2} {\tilde \partial}_{k} \!\! \int_{q} \, \bigg\{ \Gamma^{{\tilde \phi}{\tilde \phi}{\tilde \phi}\phi}_{k, a b c d}
(p,-p,q) \, G^{\rmR}_{k, d c} (q) \\
&& \hspace{24pt} +\: \Gamma^{{\tilde \phi}{\tilde \phi}\phi{\tilde \phi}}_{k,a b c d} (p,-p,q) \, G^{\rmA}_{k, d c} (q) \\
&& \hspace{24pt} +\: \Gamma^{{\tilde \phi}{\tilde \phi}\phi\phi}_{k, a b c d} (p,-p,q) \, i F_{k, d c} (q) \bigg\} ~.}
The spectral self-energy satisfies the flow equation
\eqna{
&&
\begin{picture}(249,14) (6,-45)
    \SetWidth{0.5}
    \SetColor{Black}
    \Text(8,-47)[lb]{$\displaystyle {\dot \Sigma}^{\rho}_{k, a b} \,\,=\,\, {\dot \Gamma}^{\phi{\tilde \phi}}_{k,ab} \, - 
    \, {\dot \Gamma}^{{\tilde \phi}\phi}_{k,ab}$}
\end{picture} \\
&&
\begin{picture}(249,34) (29,-52)
\SetWidth{0.5}
    \SetColor{Black}
    \Text(137,-42)[lb]{$\displaystyle +$}
    \Text(187,-42)[lb]{$\displaystyle +$}
    \Text(40,-48)[lb]{$\,\, =\,\, - \frac{i}{2} \,\, {\tilde \partial}_k \,\, \bigg\{$}
  \Text(105,-56)[lb]{$a$}
  \Text(122,-56)[lb]{$b$}
  \Text(155,-56)[lb]{$a$}
  \Text(172,-56)[lb]{$b$}
  \Text(205,-56)[lb]{$a$}
  \Text(222,-56)[lb]{$b$}
    \SetWidth{1.0}
    \DashCArc(115,-34)(10,90,-90){2}
    \CArc(115,-34)(10,-90,90)
    \SetWidth{0.5}
    \Vertex(115,-44){3.4}
    \SetWidth{1.0}
    \Line(105,-45)(115,-45)
    \DashLine(115,-45)(125,-45){2}
    \CArc(165,-34)(10,90,270)
    \DashCArc(165,-34)(10,-90,90){2}
   \SetWidth{0.5}
    \Vertex(165,-44){3.4}
    \SetWidth{1.0}
    \Line(155,-45)(165,-45)
    \DashLine(165,-45)(175,-45){2}
   \SetWidth{1.0}
    \CArc(215,-34)(10,90,-90)
    \CArc(215,-34)(10,-90,90)
    \SetWidth{0.5}
    \Vertex(215,-44){3.4}
    \SetWidth{1.0}
    \Line(205,-45)(215,-45)
    \DashLine(215,-45)(225,-45){2}
\end{picture} \\
&&
\begin{picture}(249,34) (29,-51)
\SetWidth{0.5}
    \SetColor{Black}
    \Text(90,-42)[lb]{$\displaystyle -$}
    \Text(137,-42)[lb]{$\displaystyle -$}
    \Text(187,-42)[lb]{$\displaystyle -$}
    \Text(232,-48)[lb]{$\,\, \bigg\}~.$}
  \Text(105,-56)[lb]{$a$}
  \Text(122,-56)[lb]{$b$}
  \Text(155,-56)[lb]{$a$}
  \Text(172,-56)[lb]{$b$}
  \Text(205,-56)[lb]{$a$}
  \Text(222,-56)[lb]{$b$}
    \SetWidth{1.0}
    \DashCArc(115,-34)(10,90,-90){2}
    \CArc(115,-34)(10,-90,90)
    \SetWidth{0.5}
    \Vertex(115,-44){3.4}
    \SetWidth{1.0}
    \Line(115,-45)(125,-45)
    \DashLine(105,-45)(115,-45){2}
    \CArc(165,-34)(10,90,270)
    \DashCArc(165,-34)(10,-90,90){2}
   \SetWidth{0.5}
    \Vertex(165,-44){3.4}
    \SetWidth{1.0}
    \Line(165,-45)(175,-45)
    \DashLine(155,-45)(165,-45){2}
   \SetWidth{1.0}
    \CArc(215,-34)(10,90,-90)
    \CArc(215,-34)(10,-90,90)
    \SetWidth{0.5}
    \Vertex(215,-44){3.4}
    \SetWidth{1.0}
    \Line(215,-45)(225,-45)
    \DashLine(205,-45)(215,-45){2}
\end{picture}}
\vspace*{1pt}

\subsection{Resummed $1/N$-expansion}
\vskip\pmsk

In general, flow equations for $n$-point functions are expressed in terms of $(n+2)$-point functions, and to find approximate solutions one has to truncate the infinite hierarchy of coupled flow equations. Here, we use an \emph{ansatz} for the four-point functions, which is equivalent to a resummed large-$N$ expansion of the 2PI effective action to next-to-leading order (NLO) \cite{Berges:2001fi}.

Using $O(N)$ symmetry, we can always write the two-point functions as, for instance, 
\eqn{\Gamma^{{\tilde \phi}\phi}_{k, a b} (x,y) = \Gamma^{{\tilde \phi}\phi}_{k} (x,y) \delta_{a b} ~.}
The four-point functions can be decomposed into the contributions from the different channels, e.g.\, for $\Gamma_{k}^{{\tilde \phi}\phi\phi\phi}$, we have \cite{Berges:2008sr}
\eqna{&& \hspace{-18pt}\Gamma^{{\tilde \phi}\phi\phi\phi}_{k, a b c d} (x,y,z,w) \\
&=& \Gamma^{{\tilde \phi}\phi, \phi\phi}_{k} (x,z) \delta^{(d+1)}(x-y) \delta^{(d+1)}(z-w) \delta_{a b} \delta_{c d} \\
&& +\: \Gamma^{{\tilde \phi}\phi, \phi\phi}_{k} (z,y) \delta^{(d+1)}(z-x) \delta^{(d+1)}(y-w) \delta_{a c} \delta_{b d} \\
&& +\: \Gamma^{\phi\phi , {\tilde \phi}\phi}_{k} (y,z) \delta^{(d+1)}(y-z) \delta^{(d+1)}(x-w) \delta_{b c} \delta_{a d} ~.}
In the resummed large-$N$ expansion we write these individual contributions diagrammatically as
\eqna{
&&
\begin{picture}(-28,24) (130,-238)
    \SetWidth{0.5}
    \SetColor{Black}
 \Text(52,-232)[lb]{$\Gamma_k^{\tilde{\phi}\phi , \phi \phi}(x,y)\,=\,$}
 \Text(117,-228)[lb]{$x$}
 \Text(158,-229)[lb]{$y$}
 \Text(166,-232)[lb]{~\,,}
   \SetWidth{1}
    \GOval(140,-224)(3,10)(0){0}
  \DashLine(129,-224)(121,-214){2}
  \Line(129,-224)(121,-234)
  \Line(151,-224)(159,-214)
  \Line(151,-224)(159,-234)
\end{picture} \\
&&
\begin{picture}(-28,24) (130,-234)
    \SetWidth{0.5}
    \SetColor{Black}
 \Text(52,-232)[lb]{$\Gamma_k^{\tilde{\phi}\phi , \phi \phi}(x,y)\,=\,$}
 \Text(117,-228)[lb]{$x$}
 \Text(158,-229)[lb]{$y$}
 \Text(166,-232)[lb]{~\,,}
   \SetWidth{1}
    \GOval(140,-224)(3,10)(0){0}
  \DashLine(129,-224)(121,-214){2}
  \DashLine(129,-224)(121,-234){2}
  \DashLine(151,-224)(159,-214){2}
  \Line(151,-224)(159,-234)
\end{picture}}
where the full blobs denote the resummed bubble chain with full propagators $G^{\rmR,\rmA}, F,{\tilde F}$ inserted on the internal lines connected by the bare vertices given by
\eqn{\Gamma^{{\tilde \phi}\phi\phi\phi}_{\Lambda} = - \frac{\lambda}{3N} ~, \quad \Gamma^{{\tilde \phi}{\tilde \phi}{\tilde \phi}\phi}_{\Lambda} = - \frac{\lambda}{12 N}~.}
For instance,
\eqna{
&&  
\begin{picture}(73,26) (80,-239)
  \SetWidth{0.5}
   \SetColor{Black}
 \Text(64,-227)[lb]{$\displaystyle = $}
 \Text(99,-229)[lb]{$\displaystyle - \,\, iN $}
  \SetWidth{1}
    \GOval(45,-224)(3,6)(0){0}
  \Line(38,-224)(30,-216)
  \Line(38,-224)(30,-232)
  \DashLine(52,-224)(60,-216){2}
  \Line(52,-224)(60,-232)
  \SetWidth{1}
  \Line(85,-224)(77,-216)
  \Line(85,-224)(77,-232)
  \DashLine(85,-224)(93,-216){2}
  \Line(85,-224)(93,-232)
  \SetWidth{1}
   \Line(132,-224)(124,-216)
   \Line(132,-224)(124,-232)
   \CArc(142,-224)(10,180,90)
   \DashCArc(142,-224)(10,90,180){2}
   \DashLine(152,-224)(160,-216){2}
   \Line(152,-224)(160,-232)
\end{picture}\\
&&
\begin{picture}(73,28) (175,-240)
  \SetWidth{0.5}
    \SetColor{Black}
 \Text(166,-229)[lb]{$\displaystyle + \,\, (-iN)^2 $}
 \Text(271,-228)[lb]{$\displaystyle + \,\, ... $}
  \SetWidth{1}
   \Line(215,-224)(207,-214)
   \Line(215,-224)(207,-234)
   \CArc(225,-224)(10,180,90)
   \DashCArc(225,-224)(10,90,180){2}
   \CArc(246,-224)(10,180,90)
   \DashCArc(246,-224)(10,90,180){2}
   \DashLine(256,-224)(264,-216){2}
   \Line(256,-224)(264,-232)
\end{picture}\\
&&
\begin{picture}(73,26) (80,-239)
\SetWidth{0.5}
    \SetColor{Black}
\Text(64,-227)[lb]{$\displaystyle = $}
\Text(99,-229)[lb]{$\displaystyle - \,\, iN $}
\Text(205,-229)[lb]{$\displaystyle .$}
  \SetWidth{1}
  \Line(85,-224)(77,-216)
  \Line(85,-224)(77,-232)
  \DashLine(85,-224)(93,-216){2}
  \Line(85,-224)(93,-232)
  \SetWidth{1}
   \GOval(145,-224)(3,6)(0){0}
   \Line(139,-224)(131,-216)
   \Line(139,-224)(131,-232)
  \CArc(163,-224)(10,180,90)
  \DashCArc(163,-224)(10,90,180){2}
   \GOval(181,-224)(3,6)(0){0}
  \Line(188,-224)(196,-234)
  \DashLine(188,-224)(196,-214){2}
\end{picture}}
Each vertex contributes a factor $1/N$ and each closed loop gives a factor of $N = \delta_{a b} \delta_{b a}$. Thus, all shown diagrams contribute at the same order, and there are no other diagrams that contribute at this order in the expansion. Notably, all $n$-point functions $\Gamma_{k}^{(n)}$ with $n > 4$ are of higher order in the large-$N$ expansion and the infinite hierarchy of flow equations is closed at the level of the four-point functions \cite{Gasenzer:2008zz,Berges:2008sr}. Diagrammatically this can be expressed in terms of the flow equation
\eqn{
\begin{picture}(73,26) (90,-239)
\SetWidth{0.5}
    \SetColor{Black}
\Text(82,-227)[lb]{$\displaystyle = $}
\Text(98,-229)[lb]{$\displaystyle - \,\, i N \,\, {\tilde \partial}_{k} \,\,$}
\Text(205,-228)[lb]{$\displaystyle ,$}
  \SetWidth{1}
  \Vertex(55,-214){0.8}
    \GOval(55,-224)(3,6)(0){0}
  \Line(48,-224)(40,-216)
  \Line(48,-224)(40,-232)
  \DashLine(62,-224)(70,-216){2}
  \Line(62,-224)(70,-232)
  \SetWidth{1}
   \GOval(151,-224)(3,6)(0){0}
   \Line(144,-224)(136,-216)
   \Line(144,-224)(136,-232)
  \CArc(168,-224)(10,180,90)
  \DashCArc(168,-224)(10,90,180){2}
   \GOval(186,-224)(3,6)(0){0}
  \Line(193,-224)(201,-234)
  \DashLine(193,-224)(201,-214){2}
\end{picture}}
which has only the four-vertex $\Gamma^{(4)}_{k}$ appearing on the r.h.s.\
We also note that the scale derivative represents a total derivative in this approximation \cite{Gasenzer:2008zz,Berges:2008sr,Blaizot:2010zx}, which can directly be understood from our derivation of the exact flow equation with the help of the 2PI effective action in section \ref{sec:NEQFRG}. 

Similarly, one obtains for the other four-vertices
\eqna{
&&  
\begin{picture}(73,26) (80,-239)
 \SetWidth{0.5}
    \SetColor{Black}
 \Text(64,-227)[lb]{$\displaystyle = $}
 \Text(99,-229)[lb]{$\displaystyle - \,\, iN $}
  \SetWidth{1}
    \GOval(45,-224)(3,6)(0){0}
  \DashLine(38,-224)(30,-216){2}
  \DashLine(38,-224)(30,-232){2}
  \DashLine(52,-224)(60,-216){2}
  \Line(52,-224)(60,-232)
  \SetWidth{1}
  \DashLine(85,-224)(77,-216){2}
  \DashLine(85,-224)(77,-232){2}
  \DashLine(85,-224)(93,-216){2}
  \Line(85,-224)(93,-232)
  \SetWidth{1}
   \DashLine(132,-224)(124,-216){2}
   \DashLine(132,-224)(124,-232){2}
   \CArc(142,-224)(10,180,90)
   \DashCArc(142,-224)(10,90,180){2}
   \DashLine(152,-224)(160,-216){2}
   \Line(152,-224)(160,-232)
\end{picture}\\
&&  
\begin{picture}(73,28) (175,-240)
  \SetWidth{0.5}
    \SetColor{Black}
 \Text(166,-229)[lb]{$\displaystyle + \,\, (-iN)^2 $}
 \Text(271,-228)[lb]{$\displaystyle + \,\, ... $}
  \SetWidth{1}
   \DashLine(217,-224)(209,-214){2}
   \DashLine(217,-224)(209,-234){2}
   \CArc(227,-224)(10,180,90)
   \DashCArc(227,-224)(10,90,180){2}
   \CArc(248,-224)(10,180,90)
   \DashCArc(248,-224)(10,90,180){2}
   \DashLine(258,-224)(266,-216){2}
   \Line(258,-224)(266,-232)
\end{picture}\\
&&
\begin{picture}(73,26) (80,-239)
\SetWidth{0.5}
    \SetColor{Black}
\Text(64,-227)[lb]{$\displaystyle = $}
\Text(99,-229)[lb]{$\displaystyle - \,\, iN $}
\Text(205,-228)[lb]{$\displaystyle ,$}
  \SetWidth{1}
  \DashLine(85,-224)(77,-216){2}
  \DashLine(85,-224)(77,-232){2}
  \DashLine(85,-224)(93,-216){2}
  \Line(85,-224)(93,-232)
  \SetWidth{1}
   \GOval(145,-224)(3,6)(0){0}
   \DashLine(139,-224)(131,-216){2}
   \DashLine(139,-224)(131,-232){2}
  \CArc(163,-224)(10,180,90)
  \DashCArc(163,-224)(10,90,180){2}
   \GOval(181,-224)(3,6)(0){0}
  \Line(188,-224)(196,-234)
  \DashLine(188,-224)(196,-214){2}
\end{picture}}
and
\eqna{
&& 
\begin{picture}(45,27) (100,-234)
    \SetColor{Black}
  \SetWidth{1}
    \GOval(38,-224)(3,6)(0){0}
  \DashLine(31,-224)(23,-216){2}
  \Line(31,-224)(23,-232)
  \DashLine(45,-224)(53,-216){2}
  \Line(45,-224)(53,-232)
\end{picture}\\
&&  
\begin{picture}(45,27) (145,-234)
\SetWidth{0.5}
    \SetColor{Black}
\Text(57,-236)[lb]{$\displaystyle = \, -\frac{i}{2}N \,\, \bigg\{$}
\Text(153,-229)[lb]{$\displaystyle + $}
\Text(210,-229)[lb]{$\displaystyle + $}
\Text(264,-236)[lb]{$\displaystyle \,\, \bigg\} \, $}
\SetWidth{1}
   \DashLine(115,-224)(105,-216){2}
   \Line(115,-224)(105,-232)
  \CArc(125,-224)(10,-90,90)
  \DashCArc(125,-224)(10,90,-90){2}
   \DashLine(135,-224)(145,-216){2}
   \Line(135,-224)(145,-232)
\SetWidth{1}
   \DashLine(173,-224)(165,-216){2}
   \Line(173,-224)(165,-232)
  \CArc(184,-224)(10,90,-90)
  \DashCArc(184,-224)(10,-90,90){2}
   \DashLine(195,-224)(203,-216){2}
   \Line(195,-224)(203,-232)
\SetWidth{1}
   \DashLine(230,-224)(222,-216){2}
   \Line(230,-224)(222,-232)
  \CArc(241,-224)(10,90,-90)
  \CArc(241,-224)(10,-90,90)
   \DashLine(252,-224)(260,-216){2}
   \Line(252,-224)(260,-232)
\end{picture}\\
&&
\begin{picture}(45,27) (139,-234)
\SetWidth{0.5}
    \SetColor{Black}
\Text(63,-236)[lb]{$\displaystyle + \, \frac{1}{2}(-iN)^2 \, \bigg\{ \,$}
\Text(185,-229)[lb]{$\displaystyle + $}
\SetWidth{1}
   \DashLine(129,-224)(121,-216){2}
   \Line(129,-224)(121,-232)
  \CArc(140,-224)(10,90,0)
  \DashCArc(140,-224)(10,0,90){2}
  \CArc(161,-224)(10,-90,90)
  \DashCArc(161,-224)(10,90,-90){2}
   \DashLine(172,-224)(180,-216){2}
   \Line(172,-224)(180,-232)
\SetWidth{1}
   \DashLine(203,-224)(195,-216){2}
   \Line(203,-224)(195,-232)
  \CArc(214,-224)(10,-90,90)
  \DashCArc(214,-224)(10,90,-90){2}
  \CArc(235,-224)(10,180,90)
  \DashCArc(235,-224)(10,90,180){2}
   \DashLine(246,-224)(254,-216){2}
   \Line(246,-224)(254,-232)
\end{picture}\\
&&  
\begin{picture}(45,27) (139,-234)
\SetWidth{0.5}
    \SetColor{Black}
\Text(110,-229)[lb]{$\displaystyle + $}
\Text(185,-229)[lb]{$\displaystyle + $}
\SetWidth{1}
   \DashLine(129,-224)(121,-216){2}
   \Line(129,-224)(121,-232)
  \CArc(140,-224)(10,90,0)
  \DashCArc(140,-224)(10,0,90){2}
  \CArc(161,-224)(10,90,-90)
  \DashCArc(161,-224)(10,-90,90){2}
   \DashLine(172,-224)(180,-216){2}
   \Line(172,-224)(180,-232)
\SetWidth{1}
   \DashLine(203,-224)(195,-216){2}
   \Line(203,-224)(195,-232)
  \CArc(214,-224)(10,90,-90)
  \DashCArc(214,-224)(10,-90,90){2}
  \CArc(235,-224)(10,180,90)
  \DashCArc(235,-224)(10,90,180){2}
   \DashLine(246,-224)(254,-216){2}
   \Line(246,-224)(254,-232)
\end{picture}\\
&&
\begin{picture}(45,27) (139,-234)
\SetWidth{0.5}
    \SetColor{Black}
\Text(110,-229)[lb]{$\displaystyle + $}
\Text(185,-229)[lb]{$\displaystyle + $}
\Text(257,-236)[lb]{$\displaystyle \, \bigg\} + ... $}
\SetWidth{1}
   \DashLine(129,-224)(121,-216){2}
   \Line(129,-224)(121,-232)
  \CArc(140,-224)(10,90,0)
  \DashCArc(140,-224)(10,0,90){2}
  \CArc(161,-224)(10,180,90)
  \CArc(161,-224)(10,90,180)
   \DashLine(172,-224)(180,-216){2}
   \Line(172,-224)(180,-232)
\SetWidth{1}
   \DashLine(203,-224)(195,-216){2}
   \Line(203,-224)(195,-232)
  \CArc(214,-224)(10,-90,90)
  \CArc(214,-224)(10,90,-90)
  \CArc(235,-224)(10,180,90)
  \DashCArc(235,-224)(10,90,180){2}
   \DashLine(246,-224)(254,-216){2}
   \Line(246,-224)(254,-232)
\end{picture}\\
&&
\begin{picture}(45,27) (142,-234)
\SetWidth{0.5}
    \SetColor{Black}
\Text(57,-236)[lb]{$\displaystyle = \, -\frac{i}{2}N \, \bigg\{ \,$}
\Text(175,-229)[lb]{$\displaystyle + $}
\SetWidth{1}
   \GOval(118,-224)(3,6)(0){0}
   \DashLine(111,-224)(103,-216){2}
   \Line(111,-224)(103,-232)
  \CArc(135,-224)(10,-90,90)
  \DashCArc(135,-224)(10,90,-90){2}
   \GOval(153,-224)(3,6)(0){0}
   \DashLine(159,-224)(167,-216){2}
   \Line(159,-224)(167,-232)
\SetWidth{1}
   \GOval(203,-224)(3,6)(0){0}
   \DashLine(196,-224)(188,-216){2}
   \Line(196,-224)(188,-232)
  \CArc(220,-224)(10,90,-90)
  \DashCArc(220,-224)(10,-90,90){2}
   \GOval(237,-224)(3,6)(0){0}
   \DashLine(244,-224)(252,-216){2}
   \Line(244,-224)(252,-232)
\end{picture}\\
&&
\begin{picture}(45,27) (140,-234)
\SetWidth{0.5}
    \SetColor{Black}
\Text(90,-229)[lb]{$\displaystyle +$}
\Text(172,-236)[lb]{$\displaystyle \, \bigg\} \,~. $}
\SetWidth{1}
   \GOval(118,-224)(3,6)(0){0}
   \DashLine(111,-224)(103,-216){2}
   \Line(111,-224)(103,-232)
  \CArc(135,-224)(10,90,-90)
  \CArc(135,-224)(10,-90,90)
   \GOval(153,-224)(3,6)(0){0}
   \DashLine(159,-224)(167,-216){2}
   \Line(159,-224)(167,-232)
\end{picture}}
The loop contributions appearing in the chain of bubbles diagrams are given by the expressions
\eqnlabel{\Pi^{\rmR, \rmA}_{k} (p) = \frac{\lambda}{3} \int_{q} F_{k} (p-q) G^{\rmR, \rmA}_{k} (q) ~,
\label{OneLoopRA}}
and
\eqnlabel{\Pi^{F}_{k} (p) = \frac{\lambda}{6} \int_{q} \Big[ F_{k} (p-q) F_{k} (q) - \frac{1}{4} \rho_{k} (p-q) \rho_{k} (q) \Big] ~.
\label{OneLoopF}} 
In terms of these one-loop expressions the four-point functions read
\eqna{
&&
\begin{picture}(73,32) (80,-236)
  \SetWidth{0.5}
    \SetColor{Black}
 \Text(65,-234)[lb]{$\displaystyle = \,\, \Gamma^{{\tilde \phi}\phi\phi\phi}_{\Lambda} \left\{ \, 1 - \Pi^{\rmA}_{k} + 
\left( \Pi^{\rmA}_{k} \right)^{2} + \ldots \, \right\}$}
  \SetWidth{1}
    \GOval(45,-224)(3,6)(0){0}
  \Line(38,-224)(30,-216)
  \Line(38,-224)(30,-232)
  \DashLine(52,-224)(60,-216){2}
  \Line(52,-224)(60,-232)
\end{picture}\\
&&
\begin{picture}(73,24) (80,-236)
  \SetWidth{0.5}
    \SetColor{Black}
 \Text(65,-234)[lb]{$\displaystyle = \,\, - \frac{\lambda}{3 N} + \frac{\lambda_{\text{eff},k}}{3 N} \Pi^{\rmA}_{k} ~,$}
\end{picture}\\
&&
\begin{picture}(73,32) (80,-238)
  \SetWidth{0.5}
    \SetColor{Black}
 \Text(65,-234)[lb]{$\displaystyle = \,\, i \frac{\lambda}{3 N} \left\{ \, 1 - \Pi^{\rmR}_{k} - \Pi^{\rmA}_{k} +
\ldots \, \right\}$} 
  \SetWidth{1}
    \GOval(45,-224)(3,6)(0){0}
  \DashLine(38,-224)(30,-216){2}
  \Line(38,-224)(30,-232)
  \DashLine(52,-224)(60,-216){2}
  \Line(52,-224)(60,-232)
\end{picture}\\
&&
\begin{picture}(73,24) (80,-236)
  \SetWidth{0.5}
    \SetColor{Black}
 \Text(65,-234)[lb]{$\displaystyle = \,\,  i \frac{\lambda_{\text{eff},k}}{3 N} \Pi^{F}_{k} ~,$}
\end{picture}\\
&&
\begin{picture}(73,32) (80,-236)
  \SetWidth{0.5}
    \SetColor{Black}
 \Text(65,-234)[lb]{$\displaystyle = \,\, \Gamma^{{\tilde \phi}{\tilde \phi}{\tilde \phi}\phi}_{\Lambda} \left\{ \, 1 - \Pi^{\rmA}_{k} + 
\left( \Pi^{\rmA}_{k} \right)^{2} + \ldots \, \right\}$}
  \SetWidth{1}
    \GOval(45,-224)(3,6)(0){0}
  \DashLine(38,-224)(30,-216){2}
  \DashLine(38,-224)(30,-232){2}
  \DashLine(52,-224)(60,-216){2}
  \Line(52,-224)(60,-232)
\end{picture}\\
&&
\begin{picture}(73,24) (80,-236)
  \SetWidth{0.5}
    \SetColor{Black}
 \Text(65,-234)[lb]{$\displaystyle = \,\, - \frac{\lambda}{12 N} + \frac{\lambda_{\text{eff},k}}{12 N} \Pi^{\rmA}_{k} ~,$}
\end{picture}
}
where we have defined the momentum-dependent effective coupling
\eqnlabel{\lambda_{\text{eff},k} (p) = \frac{\lambda}{\big[ 1 + \Pi^{\rmR}_{k} (p) \big] \big[ 1 + \Pi^{\rmA}_{k} (p) \big]} ~,
\label{EffectiveCoupling}}
to express the set of resummed diagrams. In order to observe the equivalence with the previous expressions above, we note that, e.g., for the set of diagrams contributing to the vertices $\Gamma^{\phi\phi\phi{\tilde \phi}}_{k}$ and $\Gamma^{{\tilde \phi}{\tilde \phi}{\tilde \phi}\phi}_{k}$, we have 
\eqna{&& \hspace{0pt}\frac{1}{1+\Pi^{\rmR}_{k} (p)} \frac{1}{1+\Pi^{\rmA}_{k} (p)} \\
&& \hspace{-2pt} =\: 1 - \Pi^{\rmR}_{k} (p) - \Pi^{\rmA}_{k} (p) + \left( \Pi^{\rmR}_{k} (p) \right)^{2} + \left( \Pi^{\rmA}_{k} (p) \right)^{2} + \ldots ~,}
where mixed products of the type  $\Pi^{\rmR}_{k} (p) \Pi^{\rmA}_{k} (p)$ vanish in the expansion, since there are no vertices $\Gamma^{{\tilde \phi}{\tilde \phi}\phi\phi}$ that would allow for that particular combination of retarded and advanced one-loop diagrams $\Pi^{\rmR,\rmA}$.

Plugging the vertices into the flow equation for the self-energies $\Sigma^{\rho,F}_{k}$ and integrating the total scale derivative gives the final result for the statistical component
\eqna{
&& \begin{picture}(3,26) (118,-232)
    \SetWidth{0.5}
    \SetColor{Black}
  \Text(26,-234)[lb]{$\displaystyle \Sigma_k^{\rho \,} \,\,=\,\, -i \,\, \bigg\{ $}
  \Text(134,-227)[lb]{$\displaystyle +$}
    \SetWidth{1}
    \GOval(102,-224)(3,10)(0){0}
  \Line(80,-224)(90,-224)
  \DashLine(114,-224)(124,-224){2}
  \CArc(102,-224)(12,0,90)
  \DashCArc(102,-224)(12,90,180){2}
      \SetWidth{1}
    \GOval(172,-224)(3,10)(0){0}
  \Line(150,-224)(160,-224)
  \DashLine(194,-224)(184,-224){2}
  \CArc(172,-224)(12,0,180)
\end{picture}
\\
&& \begin{picture}(3,26) (118,-232)
    \SetWidth{0.5}
    \SetColor{Black}
     \SetWidth{1}
  \Text(65,-227)[lb]{$\displaystyle -$}
  \Text(134,-227)[lb]{$\displaystyle -$}
  \Text(201,-232)[lb]{$\displaystyle \bigg\} ~, $}
    \GOval(102,-224)(3,10)(0){0}
  \DashLine(80,-224)(90,-224){2}
  \Line(114,-224)(124,-224)
  \DashCArc(102,-224)(12,0,90){2}
  \CArc(102,-224)(12,90,180)
  \SetWidth{1}
    \GOval(172,-224)(3,10)(0){0}
  \DashLine(150,-224)(160,-224){2}
  \Line(194,-224)(184,-224)
  \CArc(172,-224)(12,0,180)
 \end{picture}}
and the spectral component
\eqna{
&& \begin{picture}(3,26) (100,-230)
    \SetWidth{0.5}
    \SetColor{Black}
  \Text(25,-230)[lb]{$\displaystyle \Sigma_k^{F} \,\,=\,\, $}
  \Text(113,-228)[lb]{$\displaystyle +$}
    \SetWidth{1}
    \GOval(80,-224)(3,10)(0){0}
  \DashLine(58,-224)(68,-224){2}
  \DashLine(92,-224)(102,-224){2}
  \CArc(80,-224)(12,0,90)
  \DashCArc(80,-224)(12,90,180){2}
      \SetWidth{1}
    \GOval(155,-224)(3,10)(0){0}
  \DashLine(133,-224)(143,-224){2}
  \DashLine(177,-224)(167,-224){2}
  \DashCArc(155,-224)(12,0,90){2}
  \CArc(155,-224)(12,90,180)
\end{picture} \\
&& \begin{picture}(3,26) (251,-230)
    \SetWidth{0.5}
    \SetColor{Black}
  \Text(195,-228)[lb]{$\displaystyle +$}
  \Text(260,-228)[lb]{$\displaystyle .$}
  \SetWidth{1}
    \GOval(230,-224)(3,10)(0){0}
  \DashLine(208,-224)(218,-224){2}
  \DashLine(252,-224)(242,-224){2}
  \CArc(230,-224)(12,0,180)
\end{picture}
}
Writing these expression explicitly in momentum space, we have
\eqna{\Sigma^{F}_{k} (p) &=& - \frac{1}{3 N} \int_{q} \lambda_{\text{eff},k} (p-q) \left\{ \Pi^{F}_{k} (p-q) F_{k} (q) \right. \\
&& \left. -\: \frac{1}{4} \Pi^{\rho}_{k} (p-q) \rho_{k} (q) \right\} ~, \IEEEyessubnumber\label{SelfEnergy1} \\
\Sigma^{\rho}_{k} (p) &=& - \frac{1}{3 N} \int_{q} \lambda_{\text{eff},k} (p-q) \left\{ \Pi^{F}_{k} (p-q) \rho_{k} (q) \right.  \\
&& \left. +\: \Pi^{\rho}_{k} (p-q) F_{k} (p) \right\} ~, \IEEEyessubnumber\label{SelfEnergy2}}
which has the structure of a two-loop self-energy, however, with a momentum-dependent coupling $\lambda_{\text{eff},k}$.

To summarize, we have found that in the large-$N$ expansion to NLO the infinite hierarchy of flow equations is closed on the level of four-point diagrams. These can be solved directly by integrating the total scale derivative $\tilde{\partial}_{k}$. Thereby we obtain the full expressions for the self-energies and vertices to this order. In the following section we will investigate the scaling behavior of the self-energies that enter the stationarity condition \eqref{Stationarity}. That way we can extract the scaling exponent $\kappa$ characterizing different types of fixed points.

\section{Strong vs.\ weak wave turbulence}
\label{sec:nonthermalfp}

\subsection{Nonperturbative stationary transport}
\vskip\pmsk

From the integrated self-energies \eqref{SelfEnergy1} and \eqref{SelfEnergy2} we can directly classify the scaling solutions in the limit $k \rightarrow 0$, where the regulator is sent to zero. In order to better compare the nonperturbative aspects of these results to the perturbative discussion of section \ref{sec:weakturbulence}, we write without loss of generality for $k=0$:
\eqnlabel{F(p) = -i \left( n(p) + \frac{1}{2} \right) \rho(p) ~,
\label{GeneralizedFDR}}
and the presentation follows to a large extent Ref.~\cite{Berges:2010ez}.
Equivalently to \eqref{FDRCenterCoordinates}, 
the function $n(p)$ depends on the four-momentum $p = (p^{0},\mbf{p})$, in contrast to the case of thermal 
equilibrium (see section~\ref{sec:Intro}). It satisfies the symmetry property 
\eqn{n(-p) = - \left( n(p) + 1 \right)~,}
which follows from $F(-p) = F(p)$, and $\rho(-p) = - \rho(p)$.

We then write the stationarity condition \eqref{Stationarity} in terms of $n(p)$ and the spectral function $\rho(p)$ using the identity
\eqna{&& \hspace{-15pt} \Sigma^{\rho} (p) F(p) - \Sigma^{F} (p) \rho(p) \\
&=& i \left( \Sigma^{F} (p) - \frac{i}{2} \Sigma^{\rho} (p) \right) \left( F(p) + 
\frac{i}{2} \rho(p)\right) \\
&-& i \left( \Sigma^{F} (p) + \frac{i}{2} \Sigma^{\rho} (p) \right) \left( F(p) - \frac{i}{2} \rho(p)\right) ~,  \IEEEyesnumber
\label{StationarityIdentity}}
where by \eqref{GeneralizedFDR} we have
\eqna{F(p) + \frac{i}{2} \rho(p) &=& -i n(p) \rho(p) ~,\\
F(p) - \frac{i}{2} \rho(p) &=& -i \left( n(p) + 1 \right) \rho(p) ~.}
From the self-energies \eqref{SelfEnergy1} and \eqref{SelfEnergy2} we construct the linear combinations
\eqna{&&\hspace{-10pt} \Sigma^{F} (p) \mp \frac{i}{2} \Sigma^{\rho} (p) \\
&=& - \frac{\lambda}{18 N} \int_{1,2} \lambda_{\text{eff}} (p-k_{1}) \\
&& \times\: \left( F(p-k_{1}-k_{2}) \mp \frac{i}{2} \rho(p-k_{1}-k_{2})\right) \\ 
&& \times\: \left( F_{1} \mp \frac{i}{2} \rho_{1} \right) \left( F_{2} \mp \frac{i}{2} \rho_{2} \right) ~,}
that enter the stationarity equation \eqref{StationarityIdentity}. In terms of $n(p)$ this reads
\eqna{\Sigma^{F} (p) - \frac{i}{2} \Sigma^{\rho} (p) &=& - i \frac{\lambda}{18 N} \int_{1,2} \lambda_{\text{eff}} (p-k_{1}) \\
&& \times\: \left( n(p-k_{1}-k_{2}) + 1 \right) \rho(p-k_{1}-k_{2}) \\
&& \times\: \left( n_{1} + 1 \right) \rho_{1} \left( n_{2} + 1 \right) \rho_{2} ~, \\
\Sigma^{F} (p) + \frac{i}{2} \Sigma^{\rho} (p) &=& - i \frac{\lambda}{18 N} \int_{1,2} \lambda_{\text{eff}} (p-k_{1}) \\
&& \times\: n(p-k_{1}-k_{2}) \rho(p-k_{1}-k_{2}) \\
&& \times\: n_{1} \rho_{1} \, n_{2} \rho_{2} ~.}
Finally, putting everything together we obtain the following form for the stationarity condition:
\eqna{&& \hspace{-20pt} - i \left( \Sigma^{\rho} F - \Sigma^{F} \rho \right) (p) \\
&=& - \frac{\lambda}{18 N} \int_{1,2,3} (2 \pi)^{d+1} \delta^{(d+1)} (p-k_{1}-k_{2}-k_{2}) \\
&& \times\: \lambda_{\text{eff}} (p-k_{1}) \bigg\{ \left( n_{1} + 1 \right) \left( n_{2} +1\right) \left( n_{3} + 1\right) n_{p} \\
&& \hspace{52pt} -\: n_{1} n_{2} n_{3} \big( n_{p} + 1  \big) \bigg\} \, \rho_{1} \rho_{2} \rho_{3} \rho_{p} ~. \\ && \IEEEyesnumber\IEEEeqnarraynumspace
\label{Stationarity2}} 
We may bring this expression into a form which can be directly compared to kinetic or Boltzmann descriptions by mapping onto positive frequencies $p^{0}$. In particular, we consider the `collision integral'
\eqn{-i \int^{\infty}_{0} \frac{\ud p^{0}}{2 \pi} \left( \Sigma^{\rho} F - \Sigma^{F} \rho \right) (p^{0},\mbf{p}) \equiv C^{\textrm{NLO}} (\mbf{p}) ~,}
which will allow us to relate this discussion to the presentation in section \ref{sec:weakturbulence}. Here, the upper index $\textrm{NLO}$ indicates that this is accurate to next-to-leading order in the large-$N$ expansion, including processes to all orders in the coupling constant, in contrast to the perturbative discussion in section \ref{sec:weakturbulence}.

After performing the positive frequency integral, we get
\eqna{&&
C^{\textrm{NLO}}(\mbf{p}) =\\
&& \int\!\ud\Gamma_{2\leftrightarrow 2} \, \bigg\{\big( n_{p} + 1 \big) \left( n_{1} + 1 \right) n_{2} n_{3} \\
&&
-\: n_{p} n_{1} \left( n_{2} + 1 \right) \left( n_{3} + 1 \right) \bigg\} \\
&& +\: \int\!\ud\Gamma_{1\leftrightarrow 3}^{(a)} \, \bigg\{ \big( n_{p} + 1 \big) \left( n_{1} + 1 \right) \left( n_{2} + 1 \right) n_{3} \\
&& 
-\: n_{p} n_{1} n_{2} \left( n_{3} + 1 \right) \bigg\} \\
&& +\: \int\!\ud\Gamma_{1\leftrightarrow 3}^{(b)} \, \bigg\{ \big( n_{p} + 1 \big) n_{1} n_{2} n_{3} \\
&& 
-\: n_{p} \left( n_{1} + 1 \right) \left( n_{2} + 1 \right) \left( n_{3} + 1 \right) \bigg\} \\
&& +\: \int\!\ud\Gamma_{0\leftrightarrow 4} \, \bigg\{ \big( n_{p} + 1 \big) \left( n_{1} + 1 \right) \left( n_{2} + 1 \right) \left( n_{3} + 1 \right) \\
&& 
- n_{p} n_{1} n_{2} n_{3} \bigg\} ~, \IEEEyesnumber
\label{CollisionIntegral}}
with the $2^{3} = 8$ contributions from the different orthants in frequency space and
\eqna{&& \hspace{-22pt} \int\!\ud\Gamma_{2\leftrightarrow 2} \\
&=& \frac{\lambda}{18 N} \int_{1,2,3}^{(>)} (2 \pi)^{d+1} \delta^{(d+1)} ( p + k_{1} - k_{2} - k_{3}) \\
&& \times\:  \left[ \lambda_{\text{eff}} (p+k_{1}) + \lambda_{\text{eff}} (p-k_{2})+ \lambda_{\text{eff}} (p-k_{3}) \right] \\
&& \times\: \rho_{1} \rho_{2} \rho_{3} \rho_{p} ~,}
\eqna{&& \hspace{-22pt} \int\!\ud\Gamma_{1\leftrightarrow 3}^{(a)} \\
&=& \frac{\lambda}{18 N} \int_{1,2,3}^{(>)} (2 \pi)^{d+1} \delta^{(d+1)} ( p + k_{1} + k_{2} - k_{3} ) \\
&& \times\: \left[ \lambda_{\text{eff}} ( p + k_{1} ) + \lambda_{\text{eff}} ( p + k_{2} ) + \lambda_{\text{eff}} ( p - k_{3} ) \right] \\
&& \times\: \rho_{1} \rho_{2} \rho_{3} \rho_{p} ~, }
\eqna{&& \hspace{-22pt}\int\!\ud\Gamma_{1\leftrightarrow 3}^{(b)} \\
&=& \frac{\lambda}{18 N} \int_{1,2,3}^{(>)} (2 \pi)^{d+1} \delta^{(d+1)} ( p - k_{1} - k_{2} - k_{3} ) \\
&& \times\: \lambda_{\text{eff}} ( p - k_{1} ) \, \rho_{1} \rho_{2} \rho_{3} \rho_{p} ~, }
\eqna{&& \hspace{-22pt}\int\!\ud\Gamma_{0\leftrightarrow 4} \\
&=& \frac{\lambda}{18 N} \int_{1,2,3}^{(>)} (2 \pi)^{d+1} \delta^{(d+1)} ( p + k_{1} + k_{2} + k_{3} ) \\
&& \times\: \lambda_{\text{eff}} ( p + k_{1} ) \, \rho_{1} \rho_{2} \rho_{3} \rho_{p} ~.}
Here, the $(>)$ sign on the integrals indicates, that the integrals run over positive frequencies, i.e.\
\eqn{\int^{(>)}_{1,2,3} \equiv \int_{0}^{\infty} \left( \frac{\ud p^{0}}{2 \pi} \prod_{i = 1,2,3} \frac{\ud k_{i}^{0}}{2 \pi} \right) \int \left(\prod_{i = 1,2,3} \frac{\ud^{d}k_{i}}{(2 \pi)^{d}}\right) ~.}

From the Boltzmann description of the weakly interacting theory we clearly recognize from the first contribution above $2\leftrightarrow 2$ scattering processes, however, with an effective coupling $\lambda_{\text{eff}}(p)$. This momentum-dependent coupling is a consequence of the summation of an infinite number of processes, which will be crucial in order to be able to discuss the nonperturbative regime of strong turbulence at low momenta as is explained in the following. All other processes are `off-shell' and turn out not to play an important role in this context~\cite{Berges:2008wm,Berges:2008sr,Scheppach:2009wu}. 

\subsection{Scaling solutions}
\vskip\pmsk

We want to study the nonequilibrium steady states and their scaling properties. 
Along the lines of our discussion in section~\ref{sec:weakturbulence}, we can extract these properties from a scaling analysis. 
For a discussion in terms of a direct determination of the principal zeros of the collision integral \eqref{CollisionIntegral}, we refer to Refs.~\cite{Berges:2008wm,Berges:2008sr,Scheppach:2009wu}.\footnote{Ref.~\cite{Berges:2008sr} tacitly assumes the absence of particle number changing processes for the derivation of the particle cascade. For a proper discussion of this aspect see Ref.~\cite{Scheppach:2009wu}.} 

From the quantities entering the collision integral \eqref{CollisionIntegral} it is clear that we need the scaling properties of $n(p)$, the effective coupling $\lambda_{\text{eff}} (p)$, and the measure. For that purpose, we consider first the scaling behavior of $n(p)$. Extending the discussion of section~\ref{sec:weakturbulence}, we also take into account a possible dynamic critical exponent $z$ different from one and a nontrivial anomalous dimension $\eta$ following section \ref{sec:basics}. With the scaling \emph{ansatz} given by
\eqref{ScalingAnsatz1} and \eqref{ScalingAnsatz2} for the statistical and spectral function, respectively, we have
\eqn{n(p^{0} , \mbf{p}) = s^{\kappa + \eta} n(s^{z} p^{0} , s \mbf{p}) ~,}
again assuming that $n(p) \gg 1/2$. 

For the scaling analysis of the momentum-dependent effective coupling $\lambda_{\text{eff}}(p)$, we use
\eqn{ \left( \Pi^{\rmA} (p) \right)^{\ast} = \Pi^{\rmA} (-p) = \Pi^{\rmR} (p) ~,} 
to write \eqref{EffectiveCoupling} as
\eqna{\lambda_{\text{eff}} (p) &=& \frac{\lambda}{| 1 + \Pi^{\rmR} (p) |^{2}} ~. \IEEEyesnumber
\label{EffectiveCoupling2}}
To extract its scaling, we need the scaling behavior of the one-loop diagram
\eqn{\Pi^{\rmR} (p) = \frac{\lambda}{3} \int_{q} F(p-q) G^{\rmR} (q) ~, }
which is obtained from the statistical propagator $F$ and the representation of retarded propagator in terms of the spectral function $G^{\rmR}(x,y) = \rho(x,y)\theta(x^{0}-y^{0})$, i.e.\
\eqn{G^{\rmR} (p^{0},\mbf{p}) = s^{2-\eta} G^{\rmR} (s^{z} p^{0} , s \mbf{p}) ~.}
Thus, we have
\eqna{\Pi^{\rmR} (p) &=& \frac{\lambda}{3} \int_{q} F(p-q) G^{\rmR} (q) \\
&=& \frac{\lambda}{3} \int_{q} s^{2+ \kappa} F(s^{z} (p^{0} - q^{0}) , s ( \mbf{p} - \mbf{q} ) ) \\
&& 
\times\: s^{2 - \eta} G^{\rmR} (s^{z} q^{0} , s \mbf{q}) ~,}
for the one-loop diagram $\Pi^{\rmR}$. By an appropriate rescaling of the measure of this one-loop diagram, i.e., taking $q^{0} \rightarrow s^{-z} q^{0}$ and $\mbf{q} \rightarrow s^{-1} \mbf{q}$, it can be brought to the form
\eqna{\Pi^{\rmR} (p) 
&=& s^\Delta \Pi^{\rmR} (s^{z} p^{0} , s \mbf{p} ) ~, \IEEEyesnumber
\label{OneLoopDiagramScaling}}
with the scaling exponent 
\eqn{
\Delta = 4 + \kappa - \eta - z - d \,.} 
For positive $\Delta > 0$ one observes that $\Pi^{\rmR}(p) \gg 1$ for sufficiently low momenta, such that fluctuations become important in the infrared. Inserting this result into the expression \eqref{EffectiveCoupling2} for the effective coupling, we finally get
\eqn{\Delta > 0 : \quad \lambda_{\text{eff}} (p^{0}, \mbf{p} ) = s^{-2 \Delta} \lambda_{\text{eff}} (s^{z} p^{0} , s \mbf{p}) \, .}
In contrast, for the case $\Delta \leq 0$ the scaling of the effective coupling is trivial. This case will be relevant at sufficiently high momenta, where $\Pi^{\rmR}(p)$ is small. Then, from  \eqref{EffectiveCoupling2} it follows that the effective coupling is essentially equivalent to the perturbative coupling,
\eqn{\Delta \leq 0 : \quad \lambda_{\text{eff}} (p) \simeq \lambda~.} 

{\renewcommand{\arraystretch}{1.1}
\renewcommand{\tabcolsep}{15pt}
\begin{table*}[!t]
\centering
\begin{tabular}{c|c|ccl}
& Particle cascade & Energy cascade && \\[2pt]
\cline{1-3} && \\[-9pt]
Strong turbulence & $\kappa = d + 1$ & $\kappa = d + 2$ & \rdelim\}{3}{2pt}[] & \multirow{3}{*}{quartic interaction} \\[2pt]
\cline{1-3} && \\[-9pt]
\multirow{2}{*}{Weak turbulence} & $\kappa = d - \frac{5}{3}$ & $\kappa = d - \frac{4}{3}$ && \\[4pt] 
 & $\kappa = d - 2 $ & $\kappa = d -\frac{3}{2} $ && cubic interaction
\end{tabular}
\label{tab1}
\end{table*}}

It remains to determine the scaling behavior of the measure $\int \ud\Gamma$.  We consider, for instance, the $2\leftrightarrow 2$ processes:
\eqna{&&\int \ud\Gamma_{2\leftrightarrow 2} = \\ && \frac{\lambda}{18 N} \int_{0}^{\infty} \left(\frac{\ud p^{0}}{2\pi} \prod_{i = 1,2,3} \frac{\ud k_{i}^{0}}{2\pi} \right) \\
&& \times\: \int \left(\prod_{i = 1,2,3} \frac{\ud^{d}k_{i}}{(2 \pi)^{d}} \right) (2 \pi)^{d+1} \delta^{(d+1)} (p + k_{1} - k_{2} - k_{3}) \\
&& \times\: \left[ \lambda_{\text{eff}} (p + k_{1}) + \lambda_{\text{eff}} (p - k_{2}) + \lambda_{\text{eff}} (p - k_{3}) \right] \\
&& \times\: \rho_{1} \rho_{2} \rho_{3} \rho_{p} ~.}
From the scaling analysis, we obtain
\eqna{&&\hspace{-8pt} \int \ud\Gamma_{2\leftrightarrow 2}(s^{z} p^{0} , s \mbf{p},s^{z} k_{1}^{0} , s \mbf{k}_{1},s^{z} k_{2}^{0} , s \mbf{k}_{2},s^{z} k_{3}^{0} , s \mbf{k}_{3}) \\ && = s^{-4 z - 3 d + z + d - 2 \Delta + 4 (2- \eta)} \int\ud\Gamma_{2\leftrightarrow 2} (p,k_{1},k_{2},k_{3}) ~,}
where the momentum dependencies in the measure are indicated explicitly. 
With a positive exponent for the one-loop diagram \eqref{OneLoopDiagramScaling}, $\Delta > 0$, the scaling of the measure can be written as $\sim s^{-2 \kappa - 2 \eta - z}$. In contrast, 
for $\Delta \le 0$, where the coupling is given by $\lambda_{\text{eff}} \simeq \lambda$, one finds the scaling $\sim s^{-3 z - 2 d + 8 - 4 \eta}$.
The same scaling properties are also obtained for the remaining measures for the `off-shell' $1\leftrightarrow 3$ and $0\leftrightarrow 4$ processes, which may be neglected for the following discussion. 

Putting everything together, in the infrared where strong fluctuations dominate the dynamics, we obtain
\eqn{\Delta > 0 : \quad C^{\textrm{NLO}} (\mbf{p}) = s^{\kappa + \eta - z}  C^{\textrm{NLO}}(s \mbf{p}) ~.} 
In contrast, in the case where the scaling of the coupling is trivial, i.e.\ $\lambda_{\text{eff}}(p) \simeq \lambda$, relevant at high momenta we have the scaling behavior
\eqn{\Delta \leq 0 : \quad C^{\textrm{NLO}} (\mbf{p}) = s^{3 \kappa - \eta - 3 z - 2 d + 8} C^{\textrm{NLO}} (s \mbf{p}) ~.}

First, we consider the implications of the scaling behavior in the high momentum region where \mbox{$\Delta \leq 0$} such that $\lambda_{\text{eff}}(p) \simeq \lambda$. In this case we recover the perturbative discussion of section \ref{sec:weakturbulence} and the momentum integral
\eqn{\int_{0}^{k} \ud|\mbf{p}| |\mbf{p}|^{d-1} C^{\textrm{NLO}} (\mbf{p}) ~,}
describes the particle flux through a sphere of radius $k$. Performing the scaling transformation, we obtain
\eqn{\Delta \leq 0 : \quad \kappa = d + z + \frac{\eta - 8}{3}~,}
for the scaling exponent $\kappa$. Setting the anomalous dimension $\eta = 0$ and the dynamical critical exponent $z = 1$ one may immediately verify that this reproduces the same weak wave turbulence exponent $\kappa = \frac{4}{3}$ for the particle cascade in $d = 3$ spatial dimensions as in section~\ref{sec:weakturbulence}. 

In contrast, in the infrared scaling region with positive values of $\Delta$, we obtain the exponent
\eqn{\Delta > 0: \quad \kappa = d + z - \eta ~,}
for the particle flux. This scaling behavior in the nonlinear, low-momentum  regime is also known as \emph{strong turbulence}, which is displayed schematically in Fig.~\ref{fig2} in section ~\ref{sec:weakturbulence}. Of course, we can also extract the scaling behavior associated with the energy cascade.
The scaling solutions for the relativistic scalar $N$-component theory, where $z = 1$ and $\eta = 0$ are summarized in the table. The predicted values for the exponents have been tested using classical-statistical simulations on the lattice in various dimensions \cite{Berges:2008wm,Berges:2010ez,Berges:2012us}. Similar studies have also been performed for nonrelativistic scalar theories~\cite{Scheppach:2009wu,Nowak:2010tm,Nowak:2011sk,Schmidt:2012kw}, or in the context of non-abelian gauge theories \cite{Berges:2008mr,Carrington:2010sz,Berges:2012ev}.

\section{Summary and outlook}
\label{sec:summary}

In these lectures we have discussed different aspects of the nonequilibrium functional renormalization group. As an example, we considered the physics of wave turbulence in relativistic scalar $N$-component field theory. We have seen that the standard treatment based on Boltzmann equations can only describe the perturbative regime of the associated stationary transport of conserved charges. 
In contrast, the functional renormalization group can efficiently describe the perturbative regime as well as the nonperturbative physics at low momenta. 

We have approximately solved the renormalization group equations using a (2PI) resummed large-$N$ expansion to NLO. This approximation is particularly suitable, since it provides an accurate description of the physics at not too small $N$, while it is still analytically treatable. In particular, the scale derivatives are total derivatives in this approximation, which can be trivially integrated. This allowed us to efficiently discuss the close relations and differences of the renormalization group approach with kinetic theory and 2PI effective action techniques in nonequilibrium physics.    

The renormalization group provides powerful nonperturbative approximation schemes beyond a resummed large-$N$ expansion, such as derivative expansions or expansions in powers of fields with momentum-dependent vertices which we have not considered in these lectures. While this has been explored to a large extent mainly in Euclidean spacetime for physics in thermal equilibrium \cite{\RGReviews}, much less is known in the context of far-from-equilibrium problems in quantum field theory. Here the functional renormalization group may give important insights also for the nonequilibrium dynamics of, in particular, non-abelian gauge theories, where suitable large-$N$ techniques are difficult to implement.

\section*{Acknowledgments}

We thank T.\ Gasenzer, G.\ Hoffmeister, A.\ Rothkopf, C.\ Scheppach, J.\ Schmidt, D.\ Sexty and J.\ Stockemer for collaborations on related work. 


\bibliographystyle{elsarticle-num}
\bibliography{references}

\end{document}